\documentclass[english]{jfm}
\usepackage[T1]{fontenc}
\usepackage[latin9]{inputenc}
\usepackage{verbatim}
\usepackage{amstext}
\usepackage{graphicx}
\usepackage[authoryear]{natbib}
\usepackage{babel}
\begin{document}

\title[Fluctuations across a density interface]{The role of fluctuations across a stable density interface}

\author[A. Venaille, L. Gostiaux and J. Sommeria]%
{A.\ns V\ls E\ls N\ls A\ls I\ls L\ls L\ls E$^1$ \ns
L.\ns  G\ls O\ls S\ls T\ls I\ls A\ls U\ls X$^2$ \and J.\ns S\ls O\ls M\ls M\ls E\ls R\ls I\ls A$^3$}

\affiliation{$^1$ CNRS, Laboratoire de Physique, ENS de Lyon, France, $^2$ CNRS, LMFA, \'Ecole Centrale de Lyon, France, $^3$ CNRS, LEGI, Universit\'e de Grenoble, France}

\date{\today}
\maketitle

\begin{abstract}

A statistical mechanics theory for a fluid stratified in density is presented. The predicted statistical equilibrium state is the most probable outcome of turbulent stirring. It results from a competition between turbulent transport and sedimentation  of fluid particles by buoyancy effect. An approximate equipartition between kinetic and available potential energy is obtained. The slow temporal evolution of the vertical density profile is then related to the presence of irreversible mixing, which alters the global distribution of density levels.  We propose a model in which the vertical density profile evolves through a sequence of statistical equilibrium states.

The theory is then tested with laboratory experiments in a two-layer stably stratified fluid forced from below by an oscillating grid. The turbulence produced by the grid spreads upwards in the lower region and is blocked at the interface between the dense (salty) and the light (fresh) water. The interface slowly moves upward by entraining  fresh water in the turbulent region. Quantitative measurements of density fluctuations are made by planar laser induced fluorescence.

 The density fluctuations across the interface are splitted in a "wave" part and a "turbulent" part. Temporal and spatial spectra of the wave part of the density field  are well described by a previous theory due to Phillips.
We argue that statistical mechanics predictions apply for the turbulent part of the density field. Assuming a two level global density  distribution, the theory predicts a hyperbolic tangent shape for the mean vertical density profile, in agreement with experimental observations. The theory predicts that the interface thickness is inversely proportional to the Richardson number, which is also consistent with experimental observations.  
The density fluctuations obtained after removal of the wave part fit well with the statistical equilibrium theory in the interface region. However inside the mixed layer density fluctuations are instead controlled by a balance between eddy flux downward  and dissipation by cascade to small scales.
We report exponential tails for the density pdf in this region, similar to previously observed temperature pdf in  high Rayleigh number convection experiments.

\end{abstract}

\section{Introduction}
A remarkable property of strongly stratified turbulent flows is their propensity to form density staircases with relatively thin interfaces separating regions of homogenised density \citep{ruddick1989,park1994}.
%
%
Such staircases are routinely observed in the thermocline structure of the ocean.
I  that case, they often can be attributed to double diffusive convective instabilities involving temperature and salt \citep{schmitt1994}.
However,  staircases also occur in situations where the density variations are due solely to the temperature, as for instance in lakes \citep{simpson1970}, and we will restrict ourself in this paper to cases without double diffusion. 
%
The emergence of sharp density interfaces involves small scale mixing process,  and yet their presence strongly affects the large scale transport of tracers such as pollutants or nutrients.
Understanding what sets the interface shape and its evolution is therefore a challenging problem.
A satisfactory model for the interface shape remains elusive. \cite{phillips,posmentier}  proposed a dynamical equation  for the mean vertical density profile by modelling small scale turbulence with an eddy diffusivity depending on the local density gradient.
The key assumption was to assume that the eddy diffusivity coefficient decreases sufficiently fast with the local density gradients above a given threshold, so that the eddy flux decreases with increasing density gradient.
This approach explains the emergence of a sharp  interface when stirring an initial stably (strongly) stratified linear density profile, but also predicts that the interface thickness becomes infinitely small (only limited by molecular diffusivity).
By contrast,  the density interfaces observed by \citep{park1994} (with a permanent stirring) were found to be larger than the diffusive length scale. 
In order to overcome this difficulty, \cite{balmforth1998} proposed a model in which the eddy diffusion of density  is coupled to a dynamical equation for the turbulent kinetic energy -- also based on eddy viscosity.
Under some assumptions on the forcing terms in the energy equation, \cite{balmforth1998} obtained solutions for the density profile showing the formation of homogeneous layers separated by sharp but finite interfaces.

Here we propose a radically different, but complementary  approach:  the emergence of a sharp but finite interface is interpreted as the most probable outcome of turbulent stirring between two regions of homogenised density, just as for the vorticity field in freely evolving two-dimensional turbulence  \citep{miller1990,robert1991}.
We show that the  most probable state results from a competition between turbulent transport that tends to increase the interface thickness, and buoyancy effects that tend to sharpen the interface by the vertical drift of fluid elements.
This approach does not require any eddy diffusivity hypothesis: it relies on the assumption that the system  sufficiently explores the phase space available under the energy conservation constraint.
We will build upon the work of \cite{tabak2004,venaillesommeria2011}, which were to our knowledge the only attempts to generalise the equilibrium statistical mechanics approach  to  stratified turbulence. 
%
%


We will use this statistical mechanics theory to interpret experimental observations of density fluctuations across a turbulent density interface. 
We will consider for that purpose an experimental idealisation of a mixed layer originally  proposed by \cite{rouse_dodu}.
An oscillating grid is localised at the bottom of a tank initially filled with a two layer stratified fluid. 
The grid-generated turbulence spreads in the homogeneous bottom layer and is blocked at the interface, where lighter fluid is slowly entrained in the turbulent layer.
Since stirring is present only on the lower part, entrainment induces a slow depletion of the layer at rest and a progression of the mixed layer thickness. 
Here we distinguish two questions:  i/ what sets the interface shape  ? ii/  what sets the entrainment across the interface ? 

Most previous studies have been focused  on the second question, see e.g. \cite{linden1979,fernando91} for a review.
 \cite{turner68} addressed the role of molecular diffusivity on the entrainment velocity.  
Here we will assume that both Reynolds number and the Peclet numbers are very large. 
The key parameter of the problem is then the Richardson number $Ri= g' L_t /e_c$, where $L_t$ and $e_c$ are the integral length scale and kinetic energy of the turbulent velocity field, and $g' $ the reduced gravity at the interface. 
Depending on this  Richardson number, different flow regime have been identified close to the interface, see e.g.  \cite{fernando_hunt2} and references therein. 
Each of these regimes has led to a different entrainment model: 
i/ At moderate Richardson numbers, entrainment is mostly due to coherent vortices of the mixed layer impinging on the density interface \citep{linden1973,sullivan1972}.
ii/ At larger Richardson number, entrainment is controlled by the breaking of Kelvin-Helmholtz instabilities  \citep{mory1991}.
iii/ At even larger Richardson number, entrainment is dominated by intermittent interfacial wave breaking \citep{fernando_hunt1}.

By contrast with these studies, the equilibrium statistical mechanics approach does not rely on a particular physical mechanism, but rather on the assumption that the flow sufficiently explores  the phase space. 
We will show that this theory provides predictions for the interface shape, but that it must be supplemented by a model for irreversible mixing  in order to describe  entrainment across the interface.  
In real flows, there may exist dynamical regimes preventing the relaxation toward an equilibrium state, but the statistical mechanics approach provides at least an interesting  limit case that has yet not been explored.

There are only few experimental and theoretical studies dealing with the vertical density profile across the interface, which is most often taken as given for theories on the entrainment velocity. In addition, most studies are limited to the estimate of the interface thickness. 
\cite{crapper1974,fernando1985} found that the interface thickness did not vary significantly with the Richardson number, but  \cite{hopfinger_toly,hannoun88}   observed that the interface thickness decreases with increasing Richardson number. \cite{E_hopfinger86} found a similar result, but with a finite asymptotic thickness depending  on the Peclet number and the Reynolds number. 
Here we present novel measurements of density fluctuations across the interface in these experiments, using planar laser induced fluorescence techniques, with index matching.
This technique was already used by \cite{hannoun88,fernando_hunt2}  in a similar experimental setting, or by \cite{guyez_flor} in the case of two layers stably stratified Taylor-Couette experiments, but these studies did not focus on the role of density fluctuations.
As pointed out by \cite{hannoun88} , the density field  of a snapshot and of a time (or spatial) averaged of the density field may be very different. We will address this important issue. 
In particular, we will show that the density interface can be decomposed into a large scale wave motion whose amplitude is well described by a theory due to \cite{phillips77}, and that removing this wave motion allows us to recover the turbulent density field predicted by statistical mechanics. 
\cite{hannoun88} already gave experimental evidence for  Phillips theory concerning the interfacial wave amplitude. Here we confirm these observations and provide additional support for the theory. We note however that the  Phillips approach is limited because it assumes that the velocity close to the interface is entirely due to the interface motion, while turbulent velocities  in the mixed layer are in reality of the same order (or even larger) and may play an important role \citep{fernando_hunt2}.
We will also  show that the mean density profile which is \textit{a priori} prescribed in the Phillips theory can be accounted for by the statistical mechanics theory.

Importantly, the experiments presented in this paper are performed with a permanent forcing mechanism, which allows us to sustain a quasi-stationary turbulent field. 
We will interpret this turbulent field as an effective heat bath that sets the level of fluctuation for the density field.
We will neglect the direct effect of buoyancy on the turbulent eddies.
Although this may be true sufficiently far form the density interface, the interplay between turbulence and stratification may be important in real flows  \citep{hopfinger_toly,mcdougall1979}.
In addition, we consider an experimental setting with only one turbulent layer. However, previous work do not show qualitative differences between the single stirred and the double stirred case \citep{fernando_hunt2}.
Note that the evolution of the vertical density profiles  in the presence of a permanent forcing differs from the studies by \cite{linden1980,whitehead2007}, who considered the evolution of the mean vertical density profile between isolated mixing events, obtained either by dropping an horizontal grid in the flow, or by moving horizontally a vertical rod, and waiting for the turbulence to decay between each mixing event.\\

The paper is organised as follows. The equilibrium statistical mechanics theory is introduced and discussed in the second section. It is shown in a third section that one must add to this equilibrium theory a model for irreversible mixing in order to describe entrainment across the interface. 
The experimental setting is presented in a fourth section. 
The fifth section contains experimental observations on interfacial waves, and statistical mechanics predictions for the mean vertical density profile. 
It is also shown in this section that the density fluctuations far from the interface are intrinsically out of equilibrium, and their pdf display exponential tails. 
We conclude and summarise the main results in the sixth section. 

\section{Turbulent density interfaces as a statistical equilibrium state}\label{sec:statmech}
%

\subsection{Boussinesq equations and their conservation laws}

Consider a flow in the Boussinesq approximation, taking place in a domain $\mathcal{V}$.
At each time $t$ and each point $\mathbf{x}=(x,y,z)\in \mathcal{V} $ the system is described by the reduced density $b=g \left(\rho-\rho_{0}\right)/\rho_{0}$ and the velocity field $\mathbf{u}$.
Here $\rho$ is the fluid density, $g$ the gravity and $\rho_0$ a reference density.
Note that  with our convention the density $b$ is the opposite of the buoyancy, and this field will be simply referred to as the density field in the remaining of this paper.
 The velocity field is non-divergent:
\begin{equation}
\nabla \mathbf{u}=0.
\end{equation} 
The flow dynamics is expressed by the advection and molecular diffusion of density $b$
\begin{equation}
\partial_{t} b +\mathbf{u}\nabla b =\kappa\Delta b,
\end{equation}
and by the momentum equation 
\begin{equation}
\partial_{t}\mathbf{u}+\mathbf{u}\nabla\mathbf{u}=-\nabla P-b\mathbf{e}_z+\nu\Delta\mathbf{u}+\mathbf{F} ,
\end{equation}
where $\mathbf{F}$ is a mechanical forcing and $k,l$ is the
vertical unit vector.


We consider here a case without forcing and dissipation  ($\mathbf{F}=0$, $\kappa=\nu=0$).
If  the velocity field remains differentiable within the domain $\mathcal{V}$ where the flow takes place, the total energy of the flow 
\begin{equation}
E=\int_{\mathcal{V}}\left(\frac{1}{2}\mathbf{u}^{2}+bz\right) \mathrm{d}\mathbf{x} \label{eq:ener}
\end{equation}
is a dynamical invariant.
In addition, as a consequence of the density advection equation, the global distribution (i.e. histogram) of density levels 
\begin{equation}
g(\sigma)=\frac{1}{|\mathcal{V}|}\int_{\mathcal{V}}\delta(b-\sigma)\mathrm{d} \mathbf{x} \label{eq:histo}
\end{equation}
is also conserved. 
Note that $g(\sigma)$ can be related to the \textit{sorted density profile} $b_s (z)$, obtained by allowing each fluid particle to settle down to its position at rest (lower potential energy), using  $-g(b_s)db_s=Adz$, where $A$ is the horizontal domain area. Therefore the function $g$ is proportional to the inverse of the derivative of this sorted profile, $g(b_s)=-A(db_s/dz)^{-1}$. 


Boussinesq equations admit other dynamical invariants called Casimir functionals, which are related to the conservation of  Ertel potential vorticity, see e.g.
\cite{salmon1998}.
However, in this paper, we will not take into account these invariants to compute the equilibrium state.
The fact that in the absence of rotation, statistical equilibria are not affected by dynamical invariants related to potential vorticity conservation has been recently discussed in detailed in the framework of the equilibrium statistical mechanics of the shallow water  system \citep{renaud2014}. 

\subsection{Equilibrium states}

Here we proceed by analogy with statistical mechanics of two dimensional Euler or quasi-geostrophic flows originally proposed by  \cite{miller1990,robert1991}, see \cite{sommeria2001,majda2006,bouchet2012,lucarini2014} for recent reviews.

We define a \emph{microscopic state} of the system  as a given fine grained density
field $b(x,y,z)$ and a non-divergent velocity field $\mathbf{u}(x,y,z)=(u,v,w)$.
These microscopic fields are relevant phase space variables because they satisfy a Liouville theorem, stating that the flow in phase space is non-divergent.
Then Liouville theorem  can be written formally as 
\begin{equation}
\int_{\mathcal{V}} \mathrm{d} x \mathrm{d} y \mathrm{d} z  \ \left( \frac{\delta \partial_t {b}}{\delta b} + \frac{\delta \partial_t {u}}{\delta u} +\frac{\delta \partial_t {v}}{\delta v} +\frac{\delta \partial_t {w}}{\delta w}  \right) = 0 \ .
\end{equation}
The proof of such a Liouville theorem for the velocity field $(u,v,w)$ is a classical result for 3D Euler dynamics.
It can for instance be obtained by decomposing the velocity field into Fourier modes, see e.g. \cite{bouchet2012} and references therein.
The generalisation to the Boussinesq system is straightforward, since the only difference with 3d Euler is the presence of an additional  term linear in $b$ in the momentum equation, as well as an additional equation describing the pure advection of density $b$ by the non-divergent velocity field.

A \emph{macroscopic state} of the system is defined by the probability density field $\rho(\mathbf{x},\sigma,\mathbf{v})$  that describes the probability to measure a given density level $b=\sigma$  and a given velocity value $\mathbf{u}=\mathbf{v}$ in the vicinity of the point $\mathbf{x}$.
The conservation laws can be expressed as  constraints on these probability field. 
Indeed, the energy defined in Eq. (\ref{eq:ener}) and the global distribution of density levels  defined in Eq. (\ref{eq:histo}) can be expressed  as a functionals of $\rho$, 
\begin{equation}
\mathcal{E}[\rho]= \int_{\mathcal{V}}   \mathrm{d}\mathbf{x} \int_{-\infty}^{+\infty}  \mathrm{d}\mathbf{v} \int_{-\infty}^{+\infty}  \mathrm{d}\sigma \ \rho\left(\frac{\mathbf{v}^{2}}{2}+\sigma z\right) \ ,\label{eq:E_rhop}
\end{equation}
\begin{equation}
\mathcal{H}_{\sigma}[\rho]=\int_{\mathcal{V}}   \mathrm{d}\mathbf{x} \int_{-\infty}^{+\infty}  \mathrm{d}\mathbf{v} \int_{-\infty}^{+\infty}  \mathrm{d}\sigma \  \rho \ .\label{eq:HHH}
\end{equation}
Finally, the probability density field $\rho$ is normalised at each point $\mathbf{x}$:
\begin{equation}
\ \int_{-\infty}^{+\infty}  \mathrm{d} \mathbf{v}\ \int_{-\infty}^{+\infty}  \mathrm{d} \sigma  \ \rho(\mathbf{x},\sigma,\mathbf{v})=1 \ . \label{eq:norm}
\end{equation} 
Just as in the case of the vorticity field in 2D turbulence, we anticipate that a typical microscopic state $b,\mathbf{u}$ picked at random for a given set of constraint  is characterised by large scale variations superimposed with wild fluctuations at small scales.
This is what motivates the introduction of the probability field $\rho$ to describe the field at a macroscopic level: the large scale flow will be obtained by computing the mean quantities 
\begin{equation}
\overline{b}(\mathbf{x}) \equiv  \int_{-\infty}^{+\infty}  \mathrm{d}\mathbf{v} \int_{-\infty}^{+\infty}  \mathrm{d}\sigma \sigma \rho(\mathbf{x},\sigma,\mathbf{v}) \ , \quad \overline{\mathbf{u}}(\mathbf{x}) \equiv  \int_{-\infty}^{+\infty}  \mathrm{d}\mathbf{v} \int_{-\infty}^{+\infty}  \mathrm{d}\sigma \mathbf{v} \rho(\mathbf{x},\sigma,\mathbf{v}) \ , \label{eq:bar_b}
\end{equation} 
while the local fluctuations around this mean flow will be statistically described  by the distribution $\rho(\mathbf{x},\sigma,\mathbf{v})$.
Classical counting arguments allow us to estimate the number of microscopic states  associated with a given macroscopic field $\rho$, and to show that the most probable state  is the one that maximises the mixing entropy 
\begin{equation}
\mathcal{S}=-\int_{\mathcal{V}}   \mathrm{d}\mathbf{x} \int_{-\infty}^{+\infty}  \mathrm{d}\mathbf{v} \int_{-\infty}^{+\infty}  \mathrm{d}\sigma \ \rho\ln\rho \ ,
\end{equation}
 among all the states that satisfy the constraints of the problem $\mathcal{E}[\rho]=E$ and $\mathcal{H}_{\sigma}[\rho]=g(\sigma)$, see e.g. \cite{miller1990,robert1991}.
This variational problem can be written in the compact form:
 \begin{equation}
S\left(E,g(\sigma)\right)=  \max_{\rho} \left\{  \mathcal{S}[\rho] \ | \ \mathcal{E}[\rho] =E ,\ \mathcal{H}_{\sigma}[\rho]=g(\sigma) \right\} \ ,  \label{eq:var_prob}
 \end{equation}
where $S\left(E,g(\sigma)\right)$ is the equilibrium entropy, and where the maximum is searched over all the density probability fields $\rho$ which are normalised at each point $\mathbf{x}$.

\subsubsection{Determination of the equilibrium}

The first step to find the equilibrium state is to compute critical points of the variational problem, i.e. to find the field $\rho$ such that first variations of the mixing entropy around this state do vanish, given the constraints of the problem. 
One needs for that purpose to introduce the Lagrange multipliers $\beta$ and $\gamma(\sigma)$ associated respectively with the energy constraint (\ref{eq:E_rhop}) and with the constraints of the global density distribution (\ref{eq:HHH}), and then compute first variations with respect to $\rho$:
\begin{equation}
 \delta\mathcal{S}-\beta\delta\mathcal{E}+\int\gamma(\sigma)\delta\mathcal{H}_{\sigma}d\sigma=0\ . \label{eq:first_var}
\end{equation}
Equation (\ref{eq:first_var}) with the normalisation constraint  Eq. (\ref{eq:norm})  yield the following necessary and sufficient condition for $\rho$ to be a critical point of the variational problem:
\begin{equation}
\rho\left(\sigma,\mathbf{x},\mathbf{v} \right)= \left(\frac{\beta}{2\pi}\right)^{3/2} e^{-\beta\frac{\mathbf{v}^2}{2} } \rho_b(z,\sigma) , \quad  \rho_b(z,\sigma)  \equiv \frac{e^{-\beta\sigma z+\gamma(\sigma)}}{\int_{-\infty}^{+\infty}  \mathrm{d} \sigma\ e^{-\beta\sigma z+\gamma(\sigma)} } \ . \label{eq:EQUILIB}
\label{eq:indep}
\end{equation}
The values of the Lagrange multipliers $\beta$ and $\gamma(\sigma)$  are (implicitly) determined by the expression of the constraints $\mathcal{E}[\rho]=E$ and $\mathcal{H}_{\sigma}[\rho]=g(\sigma)$, given by Eq. (\ref{eq:E_rhop}) and Eq.  (\ref{eq:HHH}) , respectively.
Relaxation equations towards these equilibria (for a given global distribution of density level, and for either a prescribed energy and a prescribed inverse temperature) are proposed in \cite{venaillesommeria2011}, with an application to restratification problems.
Three remarkable properties are satisfied by the equilibrium state.
 First, the probability density field (\ref{eq:EQUILIB}) is expressed as a product of the probabilities  for density $\sigma$ and velocity $\mathbf{v}$, which means that $b$ and $\mathbf{v}$ are two independent quantities at equilibrium.
Second, the predicted velocity distribution is Gaussian,  isotropic and homogeneous in space:
The local kinetic energy $e_c$ of the equilibrium state is therefore homogeneous in space, with 
\begin{equation}
e_c=\frac{1}{2}\int \mathrm{d} \sigma\  \mathrm{d} \mathbf{v} \ \mathbf{v}^{2}\rho =\frac{3}{2}\beta^{-1}.\label{eq:energy-beta}
\end{equation}
The inverse of $\beta$ defines an effective  ``temperature'' of the turbulent field.
According to Eq. (\ref{eq:energy-beta}), this temperature is  proportional to the variance of the velocity fluctuations.
Third, the distribution of density levels $\rho_b(z,\sigma)$ depends only on the vertical coordinate $z$. Using Eq. (\ref{eq:EQUILIB}) and the definition of the averaged density $\overline{b}(z)$ given in Eq. (\ref{eq:bar_b}) , the expression of the distribution of density levels can be written as 
\begin{equation}
\rho_b(\sigma,z)=\rho_b(\sigma, 0)  e^{-\beta\left(\sigma z -\int_0^z \mathrm{d} z'\ \overline{b}(z') \right)} \ , \label{eq:eq_stat} 
\end{equation}
where $\rho_b(\sigma,0)$ must be determined using the constraints  $\mathcal{H}_{\sigma}[\rho]=g(\sigma)$ given Eq. (\ref{eq:HHH}).

Any moments of the density distribution can be computed using 
\begin{equation}
\overline{b^{n}}=\int \sigma^{n}\rho d\sigma .\label{eq:eq_statmom}
\end{equation}
Applying this expression to the second moment, we can easily show from  Eq. (\ref{eq:EQUILIB}) that the density variance is proportional to the vertical gradient of the mean density, with a coefficient of proportionality given by the  inverse temperature:
\begin{equation}
\frac{\mathrm{d} \overline{b}}{\mathrm{d} z}=\beta\left(\overline{b^{2}}-\overline{b}^{2}\right).\label{eq:beta_fluct}
\end{equation}
To conclude, the kinetic energy of the equilibrium state is homogeneous in space, and the variance of the density fluctuations is proportional to this kinetic energy, with a coefficient proportional to the vertical mean density gradient. 
It is shown in Appendix A that Eq. (\ref{eq:beta_fluct}) implies equipartition between kinetic energy and available potential energy in a low energy limit, which allows to interpret the widely reported  mixing efficiency coefficient of $0.25$ as a consequence of the rapid relaxation of the system towards statistical equilibrium. 

\subsection{Sharp interfaces as a Fermi-Dirac distribution}\label{sub:twolevels}


Let us consider the particular case of an initial state composed of two density levels in equal proportion. This would for instance be the global distribution of a tank initially filled with two uniform density layers of equal thickness.
Note that one can always choose the reference density such that the two levels are symmetric around 0, writing:
\begin{equation}
\sigma  \ \in \ \left\{ -\Sigma,\ \Sigma\right\} . \label{eq:def_2lev}
\end{equation}

We introduce the probability $p$ to measure the level $\Sigma$ at  height $z$:
\begin{equation}
\rho_b = p(z) \delta \left(\sigma-\Sigma \right)+\left(1-p(z)\right) \delta \left(\sigma+\Sigma \right)
\end{equation}
According to equation (\ref{eq:EQUILIB}) and using the fact that the two density level are in equal proportions to eliminate $\gamma(\sigma)$, we obtain 
\begin{equation}
p(z)=\frac{e^{-\beta \Sigma z }}{e^{-\beta \Sigma z }+e^{\beta \Sigma z }} , \label{eq:pz}
\end{equation}
(to avoid a constant of integration, we have chosen the origin of the $z$-axis at the interface between the two layers at rest). This is reminiscent  of a Fermi-Dirac distribution. 
Indeed, for the two level system, the incompressibility constraint plays the same role as the exclusion principle for the statistics of Fermions.
Following this analogy, the density field is a collection of fluid particles with energy $e_p=\sigma z$, with a Fermi level $\varepsilon_f=0$, in thermal contact with a heat bath characterised by the inverse temperature $\beta$.
Using Eq. (\ref{eq:pz}) and  Eq.  (\ref{eq:energy-beta}), the mean density profile $\overline{b}=\Sigma p-\Sigma (1-p)$ can be expressed in term of the density jump $\Delta b =2\Sigma$ and the local kinetic energy $e_c$:
\begin{equation}
\overline{b}(z) = - \frac{\Delta b}{2}\tanh\left( \frac{z}{\Delta h} \right), \quad \Delta h \equiv \frac{4e_c}{3\Delta b}  \ , 
\label{eq:mean_reduced_density}
\end{equation}
where we have introduced the interface thickness $\Delta h$. 
This tanh profile was  previously obtained by \cite{tabak2004} using similar arguments.  However,  \cite{tabak2004} did not relate  the inverse temperature $\beta$ to an effective turbulent heat bath.\\
According to Eq. (\ref{eq:mean_reduced_density}), the Richardson number based on the interface thickness $Ri_{\Delta h}\equiv  3 \Delta h \Delta b/(4 e_c) $ is always  equal to one.
This means that the interface thickness is  the typical  vertical length of an overturning of the stable interface provided that an order one fraction of the kinetic energy inside the overturning region is transferred into potential energy.  
The interface thickness can also be expressed in term of a global Richardson number based on the layer thickness $H$. 
\begin{equation}
\Delta h = \frac{2 H}{Ri_H},  \quad  Ri_H \equiv \frac{H \Delta b}{2e_c/3} \ . \label{eq:Delta_h_H}
\end{equation}
Note that the term $3e_c/2$ is the square of a typical turbulent velocity. 
The interface is sharp when $ \Delta h \ll H $, i.e. whenever $Ri_H \gg 1$. In that case the potential energy associated with an overturning of  the density field at the domain scale $H$ is much larger than the total kinetic energy.
By contrast, the density field becomes homogeneous in the limit for which the total kinetic energy is much larger than the potential energy associated with an overturning of the density field at the domain scale $H$, in which case  $Ri_H \ll 1$.
More generally, it is clear from Eq. (\ref{eq:eq_stat}) that the mean density profile and the statistics of density levels result from a competition between turbulent transport  and buoyancy repelling. Turbulent transport is related to the turbulent kinetic energy, which increases when the inverse temperature $\beta$ decreases. Buoyancy repelling is related to the difference between values of the density levels $\sigma$ initially present in the flow. In the high energy limit ($\beta \sigma \rightarrow 0$), the density statistics are homogeneous in space, and there is no stratification. 
 
\subsection{Turbulent velocity fluctuations as a "heat bath"}

The previous results have been obtained in a microcanonical framework, assuming that the total energy is conserved.
However the total energy is not actually conserved. Given a given cut-off length scale, kinetic energy is indeed transferred to scales smaller than this cut-off, no matter how small this cut-off length scale.
This  has two consequences: i/ in the presence of viscosity, the fluctuations will be dissipated, no matter how small is the viscosity ii/ even in the absence of viscosity, the velocity field may become non-differentiable, hence breaking energy conservation. This contrasts sharply with equilibrium states of the 2D Euler equations, in which case small scale vorticity fluctuations do not contribute to the total energy, which belongs entirely to a large scale flow structure.

Therefore a forcing term is  needed in order to maintain a statistically steady state.
But in that case the system is out-of equilibrium. However, we propose here a phenomenological interpretation of the results obtained in previous subsections by assuming that the observed  density field remains close to an equilibrium state, even when the velocity field is strongly out of equilibrium. We assume  that just as in the equilibrium case above, the kinetic energy is homogeneous in space and  is related to the (dynamical) 'temperature' through Eq. (\ref{eq:energy-beta}).
This temperature is set by a balance between forcing and dissipation in the momentum equation.
Since $\beta$ is given, the relevant statistical ensemble is the canonical one, and the equilibrium state is the minimiser of the free energy 
\begin{equation}
\mathcal{F}[\rho] \equiv-\mathcal{S}[\rho]+\beta\mathcal{E}[\rho] \ .
\label{eq:FreeE}
\end{equation}
According to Eq. (\ref{eq:indep}), the total mixing entropy can be expressed as the sum of the mixing entropy associated with the density field  and the mixing entropy associated with the velocity field: 
\begin{equation}
\mathcal{S}=\mathcal{S}_b  - \frac{3}{2}  \ln \beta ,\quad \mathcal{S}_b    \equiv  - \int_{\mathcal{V}} \mathrm{d}\mathbf{x} \int_{-\infty}^{+\infty}  \mathrm{d}\sigma \ \rho_b \ln \rho_b \ , \label{eq:entro}
\end{equation}
and according to Eq. (\ref{eq:energy-beta}) the  total energy is 
\begin{equation}
\mathcal{E}[\rho_b]=\mathcal{E}_p[\rho_b]+\frac{3}{2 \beta},\quad \text{with } \mathcal{E}_p[\rho_b]= \int_{\mathcal{V}} \mathrm{d}\mathbf{x} \int_{-\infty}^{+\infty}  \mathrm{d}\sigma \ \sigma z \rho_b  \ . \label{eq:ener_ep}
\end{equation}
Since $\beta$ is given, minimising the functional $\mathcal{F}$ defined  Eq. (\ref{eq:FreeE}) is equivalent to minimising  
\begin{equation}
\mathcal{F}_b[\rho_b] \equiv -\mathcal{S}_b[\rho_b]+\beta\mathcal{E}_p[\rho_b]  \ .
\end{equation}
Finally, we obtain a variational problem in the canonical ensemble: 
\begin{equation}
F \left( \beta ,g(\sigma) \right) =  \min_{\rho_b} \left\{-\mathcal{S}_b[\rho_b]+\beta  \mathcal{E}_p[\rho_b]   \ | \ \mathcal{H}_{\sigma}[\rho_b]=g(\sigma) \right\} \ ,
 \label{eq:var_prob2}
\end{equation}
which means that we look for the probability density field $\rho_b$ that minimises a free energy while conserving the global distribution of density levels. 
To conclude, the density field characterised by its potential energy and its global distribution of density levels can be considered as a subsystem in thermal contact with an effective heat bath provided by the turbulent velocity field.
Note that  forcing is required to maintain this  turbulent velocity field,  but we still assume that no forcing and no dissipation is present in the dynamics of the  density field. 
The effect of including dissipation in the density  dynamics  is discussed in the next section.

\section{Entrainment and irreversible mixing}  \label{sub:irreversible_entrainment}
 
\begin{figure}
\begin{center}
\includegraphics[width=\textwidth]{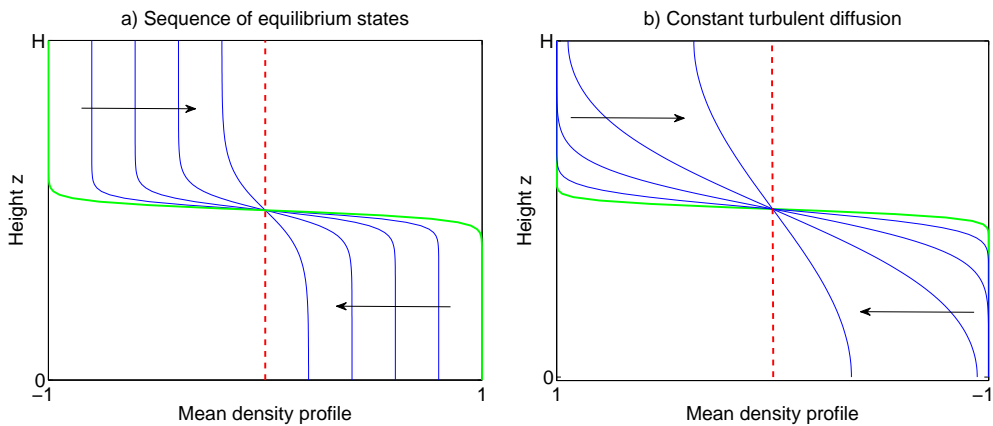}
\end{center}
\caption{a) Sequence of statistical equilibrium states in the case of a two level global density distribution ($b\in \{-\Sigma(t),\ \Sigma(t) \}$), at fixed temperature $\beta$ (prescribed by a turbulent heat bath).
The value of the density level $\Sigma(t)$ is decreasing from $1$ (plain green line) to $0$ (dashed red line). b) purely diffusive relaxation of an initial step function towards  an homogeneous density profile. The diffusion coefficient is homogeneous in space. 
 \label{fig:evolv_2lev}}
\end{figure}

In the previous section, it was assumed that the global distribution of density levels is conserved.
There is in that case no temporal evolution of the vertical mean density profile once the equilibrium state is reached.
This is because the statistical mechanics approach does not take into account irreversible mixing through turbulent cascade.
This irreversible mixing process changes the global distribution of density over time.
If this global distribution of density levels evolves on a sufficiently slow time scale, one may assume that this evolution occurs through a sequence of equilibrium states.
We show in Fig. \ref{fig:evolv_2lev}  the sequence of vertical density profiles in the case of a  two level configuration $b \in]-\Sigma,\ \Sigma[$ with decreasing values of $\Sigma$ for a fixed inverse temperature $\beta=3/(2e_c)$ (prescribed by a turbulent heat bath).
This shows a trend towards complete homogenisation of the density field with a persistence of the density interface.


 Let us assume that the density field is  anti-symmetric (in a statistical sense) with respect to an interface located at $z=0$, which is  the case for instance if the initial condition is made of two layers of homogeneous fluid with equal depth $H$.
We define the entrainment velocity as the relative temporal variation of the averaged density in the lower layer:

 \begin{equation}
U_{e}=-\frac{H}{2} \frac{\mathrm{d}_t \left<\overline{b} \right>}{ \left<\overline{b} \right>}, \quad \text{with } \left<\overline{b}  \right> \equiv \frac{1}{H}\int_{-H}^{0} \mathrm{d}z\ \overline{b} \  . \label{eq:def_Ue}
\end{equation}
This definition is consistent with Eq. (1) in \cite{turner68}.
 
 \subsection{A simplified model in the two-level case}
 
Let us now come back to the two level configuration $b \in]-\Sigma,\ \Sigma[$ in order to devise  a simple model for irreversible mixing, assuming that the dynamics goes through a sequence of equilibria.
It amounts to find a dynamical equation for the level $\Sigma(t)$.
  
 %
We propose to model the temporal evolution of the density level $\Sigma$ as a simple linear relaxation process towards the averaged density in the lower layer introduced Eq. (\ref{eq:def_Ue}):
\begin{equation}
\partial_{t}\Sigma=-s \left(\Sigma- \left<b \right>\right), \label{eq:homogenization}
\end{equation}
where $s$ is a mixing rate, i.e. the inverse of a relaxation time, which can be interpreted as a typical stretching time or a cascade rate.
The physical motivation for this model is that the interface acts as a barrier for irreversible mixing, in such a way that the turbulence tends to homogenise the fluid independently in each layer. 
Considering that $\overline{b}$ is given by the equilibrium profile Eq. (\ref{eq:mean_reduced_density}), using Eq. (\ref{eq:Delta_h_H}), and taking the limit of a sharp interface $Ri_H\gg 1$, Eq. (\ref{eq:homogenization})  becomes 
\begin{equation}
\frac{\partial_{t}\Sigma}{\Sigma}=s \frac{\log2}{Ri_H}+o\left( Ri_H^{-1} \right) \ .\label{eq:homogenization2}
\end{equation}
%
%
%
Still in the high Richardson limit ($Ri_H \gg 1$), we get  ${\partial_{t} \left<\overline{b} \right>}/{ \left<\overline{b} \right>}={\partial_{t}\Sigma }/\Sigma +o\left(Ri_H^{-1}\right)$, which, using Eq. (\ref{eq:homogenization2}) and Eq. (\ref{eq:def_Ue}), yields
\begin{equation}
 U_{e}= s \frac{H}{Ri_H} \frac{\log 2}{2}+o\left(Ri_H^{-1}\right)  \ .\label{eq:Ue_tau_mix_largeRihoverL}
\end{equation}
Assuming that the velocity field is not affected by stratification, the straining rate $s$ can be obtained on dimensional ground as  $\propto e_{c}^{1/2}/L_t$ where $e_{c}$ and $L_t$ are the turbulent kinetic energy and length scale in the absence of stratification.
We recover in that case the classical result $U_e \sim Ri_H^{-1}$ initially proposed by \cite{rouse_dodu} who assumed first  the presence of a sharp interface between two homogeneous layers, and second that the increase of the potential energy is proportional to the energy production by mechanical stirring. However, many experimental observations suggest that for very large values of $Ri_H$ the power relation between entrainment velocity and Richardson number is steeper that $-1$.
Different  arguments have been proposed to account for these observed power law, see e.g. \cite{fernando91} and references therein. 
The simplified model presented above translates the scaling for the entrainment velocity $U_e\sim Ri_H^{-n}$ into a scaling for the cascade rate $s \sim Ri_H^{-n+1}$. 
%

To conclude, we have devised a toy model in which  the mean vertical profile evolves through a sequence of equilibrium states described by a $tanh$ profile until the flow is fully homogenised.
The main caveat of the model is that it assumes a two level distribution for the global density distribution, while mixing through turbulent cascade leads to the creation of a continuum of density levels between its extremal values.
This aspect  will be discussed in more details  in the experimental part of this paper.

\subsection{Comparison with a model based on turbulent diffusion}\label{sub:erfc}

In order to appreciate the difference between the model proposed in the previous subsection and other approaches based on an effective  turbulent diffusivity, it is instructive to consider the simple case of the temporal evolution of an initial step function  through the heat equation $\partial_t{\overline{b}}=K\partial_{zz} \overline{b}$ with an homogeneous diffusion coefficient $K$ and no density flux at the upper and lower boundary $\partial_z\overline{b}|_{z=0,H}=0$. 
The sequence of vertical density profiles from the initial condition to the final homogeneous state  is shown on the right panel of Fig. \ref{fig:evolv_2lev}.  We clearly see that the route toward complete homogenisation is different in the diffusive case and in the  quasi-equilibrium case, for which the interface thickness remains quasi-constant through the homogenisation process.
%

%
Density interfaces are sometimes fitted with error functions, see e.g.  \cite{crapper1974,linden1980,whitehead2007}. The error function is the solution of the heat equation for a constant diffusion coefficient, in the case of an initial step function in an unbounded domain. In the case a bounded domain, this error function is a good fit for the density profile as long as the interface thickness remains much smaller than the domain size. 
With a proper rescaling of the  $z$ axis and of the density axis, the error function and the hyperbolic tangent functions  are hardly discernible. 
We note however that the physical mechanisms underlying the choice of one function rather than the other to fit experimental data are drastically different.
Indeed, the choice of an error function result from a model based on a local turbulent diffusivity hypothesis.
By contrast, there is no such assumption required for the choice of a tanh profile: the density profile is interpreted in that case as the equilibrium state of a two level system, which results from the competition between turbulent transport and buoyancy.

\section{Two-layer stratified fluid forced by an oscillating grid}

\subsection{Experimental setting}

\begin{figure}
\begin{center}
\includegraphics[width=\textwidth]{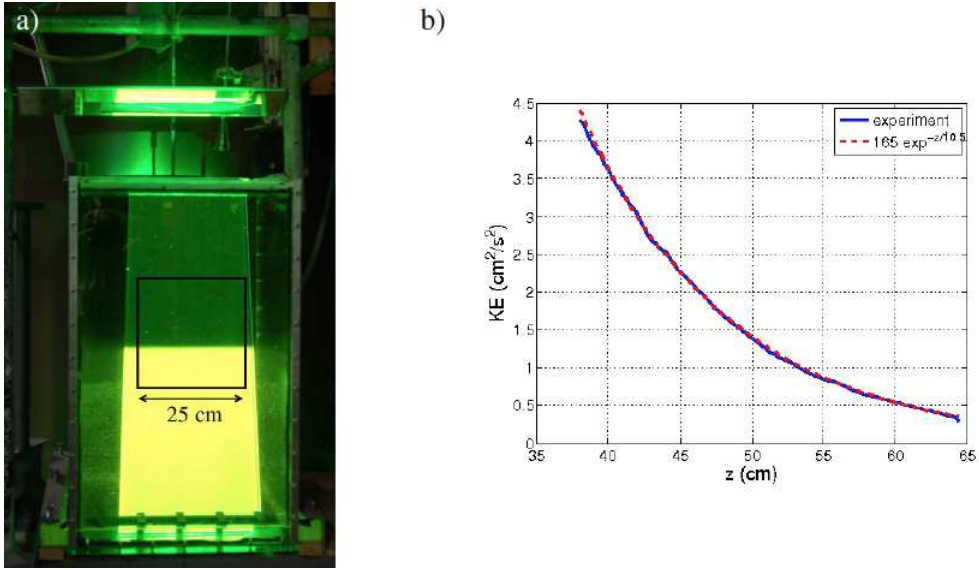} 
\end{center}
\caption{a) Experimental setting.
The tank is filled below with a layer of water with salt and rhodamine, and above with water, ethanol and rhodamine.
The LASER sheet  illuminates the centre of the tank.
 The density field is observed in the  $25\times 25 $ cm$^2$ window represented with a black line.
b) Vertical variation of the turbulent kinetic energy in the  $25\times 25 $ cm$^2$ window when there is no stratification (PIV measurements).
\label{fig:experiment}}
\end{figure}

A  tank with horizontal cross section $40 \times 40$ cm$^2$ is filled with a layer of dense fluid below a layer of light fluid, see Fig. \ref{fig:experiment}-a.
The density is homogeneous in each layer, and each layer depth is initially around $40$ cm .
This  experimental setting is similar to the one described in \cite{hopfinger_toly}.The novelty comes from measurements techniques.
We use  Planar LASER Induced Fluorescence (PLIF) with index matching between both layer in order to observe quantitatively density fluctuations.
The lower layer contains water, salt and rhodamine.
The upper layer contains water and ethanol, such that the optical index is the same in each layer.
The density difference is imposed by the concentration in salt and ethanol in both layers.

Turbulence is generated by an horizontal grid oscillating vertically at $5$ Hz, with a grid mesh of $10$ cm (including the $2$ cm thickness of the grid bars),  and an amplitude of $8$ cm (crest to crest). The same forcing is used for all the experiment.

Turbulence properties have been characterised using PIV measurements in a case without stratification.
We observed in that case an exponential decay of the kinetic energy, with an e-folding depth $L_t=10$ cm interpreted as the integral length scale  of turbulence, which is of the order of the grid mesh, see Fig.  \ref{fig:experiment}-b  and Appendix B.
We see that typical turbulent velocities close to the density interface are of the order of $U\sim 1$ cm.s$^{-1}$.
This corresponds to a Reynolds number $Re= L_t U/\nu \approx 10^3 $ associated with moderate turbulence close to the interface at the beginning of the experiment.
This means that viscous effect may be important once the interface  has moved up by around 10 $cm$. 
Note however that deeper in the mixed layer, the Reynolds number increases by two order of magnitude.
The Peclet number is $Pe=L_t U/\kappa\approx 10^6$, with $\kappa$ the salt diffusivity, also of the same order for alcohol and Rhodamine.

The main control parameter  is the density jump $\Delta \rho /\rho  $ varying from $0.01\%$ to $0.8\%$, see Tab. \ref{tab:density}.
The initial  interface height denoted $H$ is slightly different from one experiment to another.
Therefore, each experiment is characterised by two non-dimensional parameters, namely, the bulk Richardson number $Ri$ and the Richardson number based on the interface height $Ri_H$, respectively defined  by 
\begin{equation}
Ri=\frac{L_t \Delta b}{2e_t/3},\quad \text{and } Ri_H=\frac{ H \Delta b}{2e_c/3}, \label{eq:Ri}
\end{equation}
where $H$ is the initial interface height,  $e_c$ the turbulence kinetic energy measured at $z=H$ in the homogeneous case and $L_t$ the integral length scale of turbulence  in the homogeneous case.
We see Tab. \ref{tab:density} that the bulk Richardson number varies  from $1$ to $150$.
In practice, a well defined, sharp interface  was only observed for $Ri\ge 10$.
For lower bulk Richardson number, the density field did not reach a quasi-stationary state presenting a turbulent density interface.
For this reason, we will mostly focus on  experiments characterised by $Ri>10$ in order to test statistical mechanics predictions.

\begin{figure}
\begin{center}
\includegraphics[width=\textwidth]{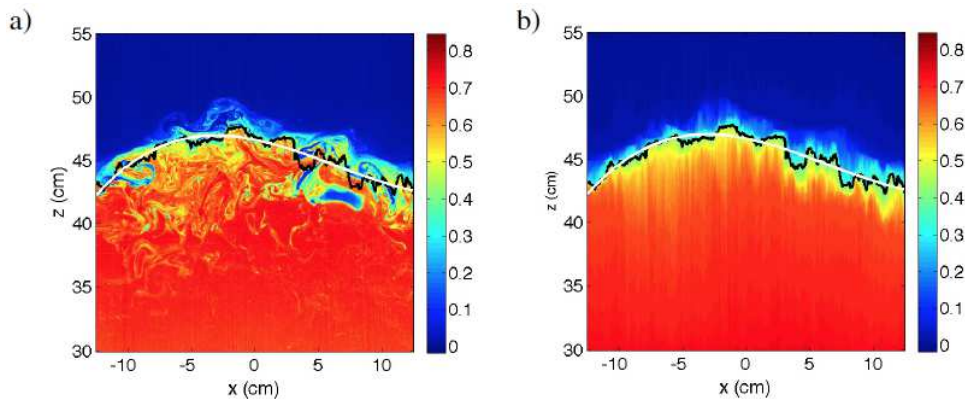}
\end{center}
\caption{a)   Snapshot of the density field at the centre of the tank, normalised between $0$ (light fluid, blue color) and  $1$ (dense fluid, red color) .
b)  Same density field, but for each value of $x$ the vertical density profile is sorted with denser fluid below, see subsection \ref{sub:def_h}. The black line is the interface defined as the height of the intermediate density  level (the contour $\alpha=1/2$) in the sorted field. The white line is a fit of this interface height with an order 2 polynomial.}
\label{fig:fields}
\end{figure}

A snapshot of the density field is shown Fig.  \ref{fig:fields}-a. 
The density field is observed in a $25\times 25$ cm$^2$ frame centred $10$ cm above the initial density interface at the beginning of the experiment, in the central part of the tank, see Fig. \ref{fig:experiment}.
For each density snapshot, light adsorption is compensated, as well as the presence of possible imperfections in the LASER sheet, as dark bands due to bubbles or dust in the optical path.\\

\begin{table}
\center
\begin{tabular}{|c|c|c|c|c|c|c|c|}
\hline
Experiment & EXP1 & EXP2 & EXP3 & EXP4 &EXP5 & EXP6 & EXP7 \\\hline
$\Delta \rho/ \rho $ ($\%$) & 0.33 & 0.3 & 0.8 & 1.3 & 0.1 & 0.4 & 0.11\\\hline
$Ri$   & 28 & 28 & 84 & 144 & 0.9 & 3.5  &  12 \\\hline
$Ri_H$ & 127 & 125 & 370 & 640 & 3.5 & 14 &  55\\\hline

\end{tabular}
\caption{Parameters for the different experiments.
The density jump $ \Delta\rho/\rho$ is  estimated as the beginning of each  experiment.
 See Eq. (\ref{eq:Ri}) for the definition of $Ri$ and $Ri_H$. \label{tab:density}} 
\end{table}

\begin{figure}
\begin{center}
\includegraphics[width=.85\textwidth]{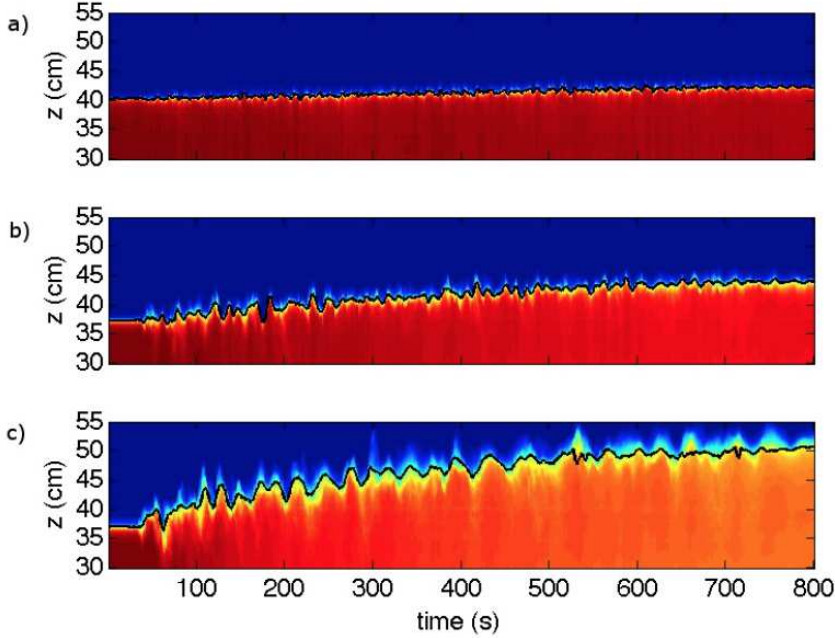}  
\end{center}
\caption{Temporal evolution of the density fields averaged in the horizontal direction for 3 different experiments, see Tab.
\ref{tab:density}: a) $\Delta b= 0.13$ m.s$^{-2}$ (EXP4) b)  $\Delta b=0.033$ m.s$^{-2}$ (EXP1)  c)  $\Delta b=0.011$  m.s$^{-2}$ (EXP7).
 The black line represents the interface (see text).
The density is normalised such that it varies between $0$ and $1$ at the initial time $t=0$ for each experiment.}
\label{fig:time_sigma_mean}
\end{figure}

Each experiment is performed during 800 second, during which a snapshot of the  density field is recorded every second.
The grid oscillation starts after $t=5$ seconds, and stopped at $t=700$ seconds. Then the relaxation to rest is observed.
The temporal evolution of the  $x$-averaged  density is shown Fig \ref{fig:time_sigma_mean} for three different experiments associated with decreasing Richardson numbers from panel a to c. It always  takes around 20 seconds before the turbulence reaches the interface.
Then the averaged density of the lower layer decreases  while the interface height increases slowly. 
In the three experiments presented in Fig.\ref{fig:time_sigma_mean}, we see qualitatively that the interface associated with the x-averaged density remains sharp, and its thickness increases with decreasing Richardson numbers. 
%
%

\subsection{Relation with the statistical mechanics model}

Several assumptions are required to interpret this  experiment in the framework of the equilibrium statistical mechanics theory introduced in section \ref{sec:statmech}.
The source of kinetic energy is localised at the grid position in the experiment, implying a vertical decay of the kinetic energy.
This contrasts with the statistical equilibrium stating that the kinetic energy (or the effective temperature) is homogeneous in space.
Our working hypothesis is that prediction from equilibrium statistical mechanics for the density field may be applied near the density interface, by considering that the effective temperature of the equilibrium state is provided by velocity fluctuations that would be observed at the interface in the absence of stratification.

A second difficulty is that there must be sufficient mixing in phase space for the system to reach the equilibrium state.
It is clear that the interface motion at large scale is dominated by waves, for which nonlinear effects driving this mixing are inhibited. 
We shall therefore assume that the turbulent fluctuations involved in the statistical equilibrium are limited to small scales, after exclusion of a "wave motion " of the density interface.
A third limitation for the applicability of the theory is the irreversible dissipation of density fluctuations through turbulent cascade, changing the global distribution of density levels with time.
As discussed in section \ref{sub:irreversible_entrainment}, this effect is actually related to entrainment across the interface.
We will assume that the global distribution of density levels evolves on a time scale longer than the one required for the system to reach the equilibrium state. 
A fourth limitation of the present experiment comes from the asymmetric forcing: only the lower layer is turbulent.
A direct consequence of this asymmetric forcing is that the turbulent layer is actually penetrating into the quiescent layer: contrary to the homogeneous case discussed in 
 previous section, the interface is shifting vertically in the experiment. 
Testing equilibrium statistical mechanics predictions requires in that case to be in the reference frame moving with the interface, assuming that the fluid evolves through a sequence of stationary states. It is thus necessary to define precisely  what is the interface height in the experiment.

\subsection{Definition of the interface height}\label{sub:def_h}

One major difficulty associated with the definition of the interface height and thickness stems from the fact that instantaneous density profiles may be very different from temporal or spatial averages,  see e.g. \cite{hannoun88}.
In order to define the interface height at each time $t$ and location $x$ , \cite{hannoun88,fernando_hunt2} considered iso-density contours parametrised by 
\begin{equation}
\alpha \equiv \frac{\rho-\rho_{min}}{\rho_{max}-\rho_{min}} \ , \label{eq:isocontours}
\end{equation}
and  defined the  interface height $h(x,t)$ as the height of the contour $\alpha=1/2$ 
As  noticed by  \cite{hannoun88,fernando_hunt2}, this method cannot be applied  when interface overturns. This occurs for instance with wave breaking. As an example,  one clearly sees on the snapshot of Fig. \ref{fig:fields}-a that iso-density contours do not  define singled-valued functions for the interface height  $h(x,t)$.

We propose and discuss in this paper a method to define the location of a corrugated interface, which is well defined even in the presence of overturning events in he density field. This method allows us to distinguish  interfacial waves form turbulent fluctuations around the interface. 
In order to find the interface height $h(x,t)$ on the density field snapshot Fig. \ref{fig:fields}-a,  a  "sorted density field" is computed Fig. \ref{fig:fields}-b:  at each horizontal point $x$, the vertical sequence of $n_z$ pixels is sorted so that the density is decreasing with increasing height $z$. 
As a consequence of this sorting procedure,  iso-density contours are always single-valued functions of the horizontal $x$ coordinate. 
The interface $h(x,t)$ is then defined as the height of the intermediate density contour $\alpha=1/2$ of the sorted density field, which is represented as a thick black line in Fig. \ref{fig:fields}-a,b.

\begin{figure}
\setlength{\unitlength}{1cm}
\begin{center}
\includegraphics[width=\textwidth]{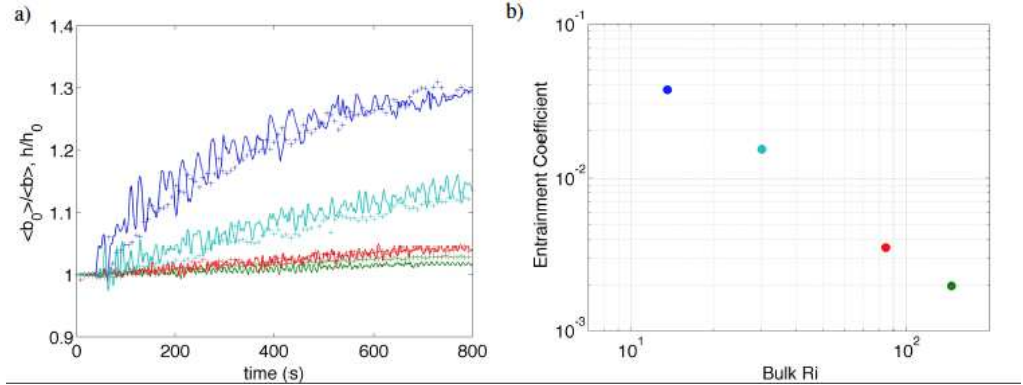}
\end{center}
\caption{ a) mass conservation and validation of PLIF measurements.
The thin plain line represents the  temporal evolution of the x-averaged interface height $\overline{h}(t)$ normalised by the initial height $h_0$, and obtained with the sorting algorithm. The "$+$" symbols represent the temporal evolution of the averaged density $<b>$ in the mixed layer, normalised by the initial density $<b_0>$ in this layer.
b)   Entrainment coefficient as a function of the average Richardson number for each experiment.
The entrainment coefficient is defined as the ratio between the interface velocity  and  the turbulent velocity.
The interface velocity is obtained by a linear fit of the curve of the left panel over the interval $t=100-300$ s.
The turbulent velocity around the interface height is  estimated by using the PIV measurements performed in the case without stratification.
}
\label{fig:entrainment}
\end{figure}

\subsection{Estimate of the entrainment velocity and test of PLIF calibration}

The spatial and temporal variability of the interface height $h(x,t)$  will be explored in  next section. 
Here we consider the slow temporal evolution of the x-averaged interface height $\overline{h}(t)$ shown Fig.  \ref{fig:entrainment}-a. 
We assume that there is a time scale separation between a fast temporal variability of the interface height and a slow  penetration of the turbulent layer into the layer at rest.  
The corresponding entrainment velocity $U_e$ is estimated by considering a linear fit of the interface height $\overline{h}(t)$ between $t=100$ and $t=300$ seconds for each experiment.
The PLIF calibration is checked  on Fig.  \ref{fig:entrainment}-a by comparing the temporal evolution of the x-averaged interface $\overline{h}(t)$ with the estimate $h_0 <b_0>/<b>$, where $<b>(t)$ is the averaged density  in the turbulent layer, below the interface, and $h_0$, $<b_0>$ are the initial interface height and density in the lowest layer. 
Despite the temporal variability of $\overline{h}(t)$ associated with interfacial waves,  there is a good  agreements between the estimate of the interface elevation  $\overline{h}(t)$ using the sorting algorithm, and the estimate of the interface elevation  $h_0 <b_0>/<b>$ using mass conservation.  
A systematic drift is observed only in the strongly stratified case, probably due to light absorption that was not completely cancelled with our data analysis procedure for this particular experiment.
The entrainment coefficient  $E=U_e/(2e_c/3)^{1/2}$ is defined as the ratio  between the entrainment velocity and the rms turbulent velocity. This rms turbulent velocity were estimated  at the interface height $z=h$,  using PIV measurements in a case without stratification. 
The variations of the entrainment  coefficient $E$ with the bulk Richardson number $Ri$ defined Eq. (\ref{eq:Ri}) is plotted Fig. \ref{fig:entrainment}-b.
We find a power law $E\sim Ri^{-n}$ with the exponent $n$ between $1$ and $3/2$, consistently with previous observations in the same range of Richardson numbers, either in similar experimental setting \citep{fernando91}, or in Taylor-Couette experiments \citep{guyez_flor}. 
In the remaining of this paper, we do not further investigate entrainment mechanism, which would require more detailed measurements of the velocity field close to the interface in order to better characterise vertical density fluxes. 
 We rather assume that the system evolves slowly through a sequence of statistically steady states, and we focus on the properties of the density fields associated with these states. 

\section{Characterisation of interfacial waves,  interface shape and density fluctuations across the interface}

On the one hand, it is clear from  Fig.  \ref{fig:fields} that the interface is corrugated and present small scale structures, consistent with the statistical mechanics approach. 
On the other hand, low frequency oscillations of the interface height are clearly visible in Fig. \ref{fig:time_sigma_mean}. These low frequency oscillations cannot be explained with the statistical mechanics approach.  
We  show in the first subsection that some properties of the observed interfacial waves are well described by a heuristic theory due to \cite{phillips77}. 
The characterisation of interfacial waves will allow us to propose a criterion to distinguish a "wave part" and a "turbulent part" for the fluctuations of the density field close to the interface. 
The "turbulent part" of the density field will be considered in the second subsection in order to test statistical mechanics predictions for the interface shape.  
Finally, we show in the last subsection that density fluctuations within the mixed layer and sufficiently far from the interface are much larger than expected from equilibrium statistical mechanics arguments, and present exponential tails.  

\subsection{Interfacial waves}

\begin{figure}
\setlength{\unitlength}{1cm}
\begin{center}
\includegraphics[width=\textwidth]{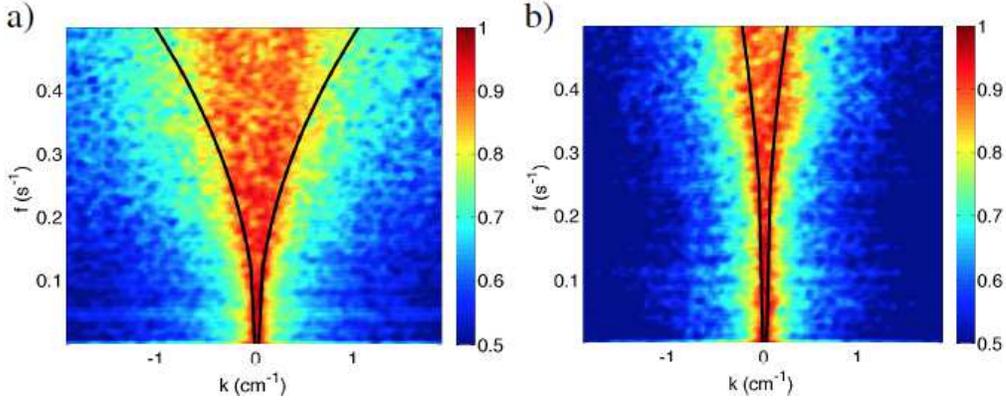}
\end{center}
\caption{ a)  Interface height spectrum for EXP1 ($\Delta b=0.03$ m.s$^{-2}$). For each frequency $f$, the spectrum is normalised by its maximal value. The black line is the dispersion relation for interfacial waves given Eq. (\ref{eq:dispersion}). b)  Same plot for EXP4 ($\Delta b=0.13$ m.s$^{-2}$).}
\label{fig:dispersion}
\end{figure}

 The spectrum of the interface elevation in a two layer fluid subject to an external forcing was predicted by  \cite{phillips77} with heuristic arguments. 
In this framework,  the predicted spectrum does not depend on the forcing mechanism. 
The only input of the theory is the interface thickness  $\Delta h$. 
We give in the following some experimental evidence for Phillips' theory, assuming that the interface thickness is the one predicted by statistical mechanics, and also discuss limitation of this approach.  

 \subsubsection{Dispersion relation for interfacial waves}
 
\cite{phillips77} considered a sharp density interface separating two homogeneous layers of height $H/2 \gg \Delta h$, with a density jump $\Delta b= g\Delta\rho/\rho$. The buoyancy frequency inside  the interface can be estimated as  
\begin{equation}
 N \sim \sqrt{\frac{\Delta b}{\Delta h}} \ . \label{eq:N} 
\end{equation}
Let us first assume that the interface elevation is a monochromatic wave characterised by the frequency $\omega$ and the wavenumber modulus $K$. 
Let us consider in addition that the wavenumber modulus is such that   $ K \ll 2\pi /(\Delta h)$, so that the interface can be considered infinitely thin at lowest order, and $K\ge 2\pi/H$, so that the limit of deep water can be considered.
Under these assumptions, the dispersion relation is 
\begin{equation}
\omega =\sqrt{\frac{\Delta b K}{2}}  \quad \text{for } N  \ll \omega \le \sqrt{\frac{\Delta b \pi }{H}} \ .\label{eq:dispersion} 
\end{equation}
%
%
%
The presence of interfacial waves is revealed  Fig. \ref{fig:dispersion} by the  spatial-temporal spectrum of the interface elevation, denoted  $\hat{\hat{h}}(f,k)$, where 
$h(x,t)$ is the observed  interface elevation  defined with the sorting algorithm introduced in subsection \ref{sub:def_h}.

For each frequency $f=\omega / 2 \pi$, the spectra shown  in Fig. \ref{fig:dispersion}  have been normalised by their maximum value over the  horizontal wavenumbers $k$. 
We see that significant contributions to the spectrum are always located inside the region delimited by the dispersion relation Eq. (\ref{eq:dispersion}).  
Note that the interface variations are measured on a line in the $x$ direction and not in the horizontal $(x,y)$ plane, so any wave numbers $|k|< 2 \omega^2/\Delta b$ may correspond to an interfacial wave, according to Eq. (\ref{eq:dispersion}).
This is why the spatio-temporal spectra  of Fig. \ref{fig:dispersion}  are not merely peaked around the dispersion relation. 

\subsubsection{Prediction of the interface elevation amplitude for a monochromatic wave}
For a given interfacial wave with wavenumber modulus $K$, the interface elevation amplitude is denoted  $a_K$. We assume that the flow around the interface is entirely due to the potential flow associated with the interface deformation. 
This allows us to estimate the velocity field close to the interface as   
\begin{equation}
U_K\sim a_K \omega \ .\label{eq:Uk} 
\end{equation} 
Given that the flow in the mixed layer is strongly turbulent, this hypothesis may be questioned, and we will provide further discussion on this point at the end of this subsection. 
%
%
When the interface is infinitely sharp ($\Delta h = 0$), the horizontal velocity field due to the variation of the interface elevation is discontinuous across the interface, with a velocity jump given by $\Delta U_K\sim a_K \omega$. 
Let us now consider that  the interface is characterised by a  small but non-zero  thickness ($\Delta h \ne 0$)
Using the estimate of the velocity jump obtained in the limit of an infinitely sharp interface, the vertical gradient of the horizontal velocity field is estimated as
\begin{equation}
\partial_z U \sim \frac{\Delta U_K}{\Delta h} \sim \frac{a_K \omega }{\Delta h} \ .  \label{eq:shear}
\end{equation}
Defining the local (or gradient) Richardson number inside the interface as
\begin{equation}
Ri_{loc}= \frac{N^2}{\left(\partial_z U\right)^2}\sim \frac{\Delta b\Delta h}{a^2_K \omega^2} \sim \frac{\Delta h}{a^2_K K} \ ,  \label{eq:Riloc}
\end{equation}
a sufficient condition for stability of the  flow  inside the thin interface is $Ri_{loc}>1/4$ \citep{miles1961}.
The key idea of \cite{phillips77} is then to assume
i/ that the flow is actually unstable whenever $Ri_{loc} <1/4$;  
ii/  that this instability eventually leads to wave breaking, which limits the growth of the wave amplitude $a_K$;
iii/ that this is the dominant mechanism to extract energy from the interfacial wave; 
iv/ that an external mechanism constantly supplies energy to the wave. 
Then a steady state can be  reached, and the saturated wave amplitude  $a_K$ is such that the condition of criticality $Ri_{loc}=1/4$ is satisfied. 
Injecting Eq. (\ref{eq:dispersion}-\ref{eq:shear}) in Eq. (\ref{eq:Riloc}), this condition for criticality yields
\begin{equation}
a^2_K  \sim \frac{\Delta h}{ K}  .\label{eq:rms_single_k}
\end{equation}
%
%
%

\subsubsection{Spatial power spectrum of the interface elevation}

Let us now assume that the interface is an (isotropic) collection of waves with wavenumber $(k,l)$ (and wavenumber modulus $K=\sqrt{k^2+l^2}$), and that these waves do not interact. 
Let  us write $\Psi(k,l)$ the spatial  power spectrum of the interface elevation. The variance of interface elevation at wave number $K$ is related to the power spectrum through 
\begin{equation}
a^2_K=\int_{\sqrt{k'^2+l'^2}>K} \mathrm{d} k' \mathrm{d} l' \ \Psi(k',l') \sim  K^2 \Psi(k,l) . \label{eq:a2_K}
\end{equation}
Injecting then  Eq. (\ref{eq:rms_single_k}) in Eq. (\ref{eq:a2_K}) yields 
\begin{equation}
\Psi(k,l)  \sim \frac{\Delta h}{ K^3} \ . \label{eq:power_spectrum_kl}
\end{equation}
The experimental spatial power spectra are obtained by measuring the interface elevation along a line in the $x$ direction. Let us call $\Psi_{x}(k)$ the  power spectrum of the interface elevation along this direction. It is given by
\begin{equation}
\Psi_{x}(k)\equiv \int_{-\infty}^{+\infty} \mathrm{d} l \ \Psi(k,l)  \sim \frac{\Delta h}{ k^2} \ . \label{eq:power_spectrum_k}
\end{equation}
The experimental observation of the interface elevation power spectrum $\Psi_x(k)/\Delta h$ is shown Fig.  \ref{fig:slosh}-a.
The spectral amplitudes have been normalised by the interface thickness predicted Eq. (\ref{eq:mean_reduced_density}) with statistical mechanics arguments, i.e. by $\Delta h \sim e_c/ \Delta b$.  
We see that the Phillips prediction of a $-2$ slope for  this spectrum  is consistent with the behaviour of the experimental spectrum at low wave numbers  in Fig.  \ref{fig:slosh}-a.
However, some care must be taken to interpret these spatial spectra for large wavenumbers. 
Indeed,  the meaning of the small scale spatial fluctuations of the interface $h(x,t)$ defined with the sorting algorithm is not clear. 
These small scales  may be dominated by the presence of turbulent fluctuations in the density field rather than by interfacial waves.
For these reasons, we will consider in the remaining the interface $h_{interp}(x,t)$, obtained for each time $t$ by fitting the interface elevation $h(x,t)$ with a third order polynomial.  
Finally, rescaling the spatial power spectrum shown Fig. \ref{fig:slosh}-a by the height $\Delta h $ allows us to obtain a reasonable  collapse of the three different experiments, consistently with Eq. (\ref{eq:power_spectrum_k}). 

\subsubsection{Temporal power spectrum of the interface elevation}

\begin{figure}
\setlength{\unitlength}{1cm}
\begin{center}
\includegraphics[width=\textwidth]{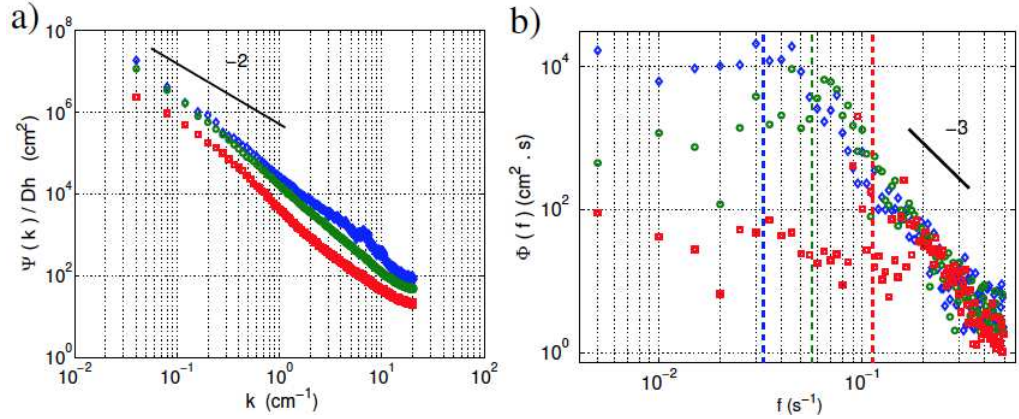}
\end{center}
\caption{a)  Spatial power spectrum. Blue diamonds:  $\Delta b=0.011$ m.s$^{-2}$ (EXP7);  Green circles:  $\Delta b=0.03$  m.s$^{-2}$ (EXP1);  Red squares:  $\Delta b=0.13$  m.s$^{-2}$ (EXP4); b) Temporal power spectra of the x-averaged interface height time series. The  vertical dashed lines represent  the location of the gravest linear interfacial mode for each experiment, given by  $\omega_0=\sqrt{\pi \Delta b / L }$, where $L$ is the lateral extension of the tank.) 
}
\label{fig:slosh}
\end{figure}

Let  us now introduce  $\Phi(\omega)$ the temporal power spectrum of the interface elevation. The variance of interface elevation at frequency  $\omega$ is 
\begin{equation}
a^2_{K}=\int_{\omega}^{+\infty} \mathrm{d} \omega' \ \Phi(\omega') \sim  \omega \Phi(\omega) . \label{eq:a2_omega}
\end{equation}
Injecting Eq. (\ref{eq:a2_omega}) in Eq. (\ref{eq:rms_single_k}), and using the dispersion relation Eq. (\ref{eq:dispersion}) yields 
\begin{equation}
\Phi(\omega) \sim \frac{\Delta b \Delta h}{ \omega^{3}} \ .
\end{equation}
Considering the statistical mechanics prediction in  Eq. (\ref{eq:mean_reduced_density})  for the interface thickness $\Delta h$,  one gets 
\begin{equation}
\Phi(\omega)  \sim \frac{e_c}{ \omega^{3}} \ . \label{eq:omega_fin}
\end{equation}
This means that the amplitude of the interface displacement frequency spectrum is independent from the density jump $\Delta b$. It only depends on the local kinetic energy. To our knowledge, this simple but important consequence of an interface thickness scaling as the inverse of the Richardson number ($Ri \sim L_t \Delta b/e_c$ or $Ri_H \sim H \Delta b/e_c$) has not been discussed previously.
Experimental observations of the interface elevation temporal power spectrum  are shown Fig. \ref{fig:slosh}-b. 
The agreement between the theoretical predictions and experimental results is  good. 
For frequency higher than the gravest linear mode of interfacial waves (represented as vertical dashed lines), the spectrum slope is consistent with the $-3$ slope predicted by Phillips theory. 
Note that the maximum observed frequencies are always larger than the buoyancy  frequency $N=\sqrt{\Delta b /\Delta h}$. 
%
Perhaps more strikingly, no rescaling have been used to plot the spectra, and yet they all collapse on the same curve in the regime where the $-3$ slope is observed.  
Since the interface elevation  is roughly similar for all the experiment , the kinetic energy $e_c$ around the interface within the mixed layer is not expected to vary significantly from one experiment to another.  
The collapse of all the experiments on the same curve confirms therefore the prediction of  Eq. (\ref{eq:omega_fin}).
This equation was  obtained under the assumption that the interface thickness varies as the inverse of the Richardson number, consistently with statistical mechanics predictions.
The collapse of the spectra on a single curve at high frequency is therefore an indirect test of these statistical mechanics prediction for the interface thickness of the mean vertical density profile.\\ 
Previous  experimental observations of a  $-3$ slope of the  temporal power spectrum of the interface elevation were provided by \cite{hannoun88}. This slope is also consistent with the experiments by \cite{fernando_hunt2}.  
However, \cite{hannoun88} found that the  amplitude of the frequency  power spectra of the interface elevation scaled as $Ri^{-1}$.  Their scaling amounted to an interface thickness decreasing as $Ri^{-2}$. By contrast, \cite{fernando_hunt2} found an interface thickness that was not varying significantly with the Richardson number. Our result is intermediate between both cases.  

Finally, we observed in the experiments at high Richardson numbers the presence of  a well identified peak in the  frequency spectrum.  The frequency of this peak were slightly smaller than the gravest linear interfacial mode represented as vertical dashed lines on Fig. \ref{fig:slosh}-b.
These sloshing frequencies may probably be attributed to nonlinear interactions between interfacial waves, and may also be associated with the presence of solitons.
 To our knowledge, there were no previous observation of such sloshing dynamics in similar experiments. Understanding this phenomenon will require further work.

\subsubsection{Consistency  of the approach}

A key assumption of Phillips theory is that close to the interface, the velocity field $U_K$ at scale $K$  is entirely due to the potential flow  created by the variations of the interface elevation. 
In the experiment, the energy source for the waves is the turbulent velocity field. 
Let us call $U_{turb,K} $ the rms turbulent velocity at scale $K$, close to the interface, within the turbulent mixed layer. This velocity contains the contribution of all wave numbers lager than $K$.
Phillips approach is consistent at scales such that $U_K \gg U_{turb,K}$.

In order to obtain a simple estimate for $U_{turb,K}$, we assume that the turbulent velocity field is unaffected by stratification, and well described by the phenomenology of three dimensional homogeneous isotropic turbulence. 
Of course, this hypothesis is too simplistic, but it can be used as a lowest order estimate. A much more detailed analysis and discussion on the coupling between turbulent and stratification is provided in  \cite{fernando_hunt1}.
For the sake of simplicity, we assume in the following that there is no turbulent motion at scales larger the integral length scale $L_t$\footnote{This assumption can not be fully valid  in our experiment, since the presence of large scale flow structure filling the whole domain are often reported in confined turbulent flows).}.
According to previous notations, the velocity field at the integral length scale is $U_{turb,K} \sim e_c^{1/2}$. 
At scales $K \gg 2\pi/L_T$, assuming an inertial range, the turbulent velocity  is obtained by dimensional analysis: $U_{turb,K} \sim \epsilon^{1/3} K^{-1/3}$, with $\epsilon \sim e_c^{3/2}/L_t$ the energy dissipation rate. 
Following \cite{carruthers1986}, we assume that temporal fluctuations of the velocity field at a given point are given by the random advection of turbulent eddies by the integral scale eddies.  This yields $\omega \sim e_c^{1/2} K$, and $U_{turb,K} \sim \epsilon^{1/3} e_c^{1/6} \omega^{-1/3}$.

Injecting Eq. (\ref{eq:omega_fin}-\ref{eq:a2_omega}-\ref{eq:dispersion}) in Eq. (\ref{eq:Uk}), we obtain the estimate $U_K\sim e_c^{1/2}$ for the velocity close to interface associated with the variation of the interface elevation. 
In other words, whatever the scale $K$ such that $2\pi /\Delta h \ll K\le 2\pi/H$, the potential flow created by the interface elevation variations is of the order of the rms turbulent velocity  $e_c^{1/2}$. 
We stress  that this result  relies on the assumption $h\sim e_c/ \Delta b $, which was done based on the statistical mechanics result Eq. (\ref{eq:mean_reduced_density}). 
We see that the condition for consistency  $U_{turb,K} \ll U_K$ is valid for sufficiently small scales  ($K\gg 2\pi/ L_t$), or for sufficiently high frequency ($\omega \gg 2 \pi e_c^{1/2}/ L_t$).  
The cut-off frequency $f_c=e_c^{1/2}/ L_t$ is of the order of $0.1$ s$^{-1}$ in the experiment. 
The maximum observed frequency is $f_{max}=1$  s$^{-1}$ on Fig. \ref{fig:slosh}-b. The criterion $f\gg f_c$ is therefore only marginally satisfied in the range of the observed $-3$ slope.
 
\subsection{The distribution of density levels}

\begin{figure}
\setlength{\unitlength}{1cm}
\begin{center}
\includegraphics[width=.9\textwidth]{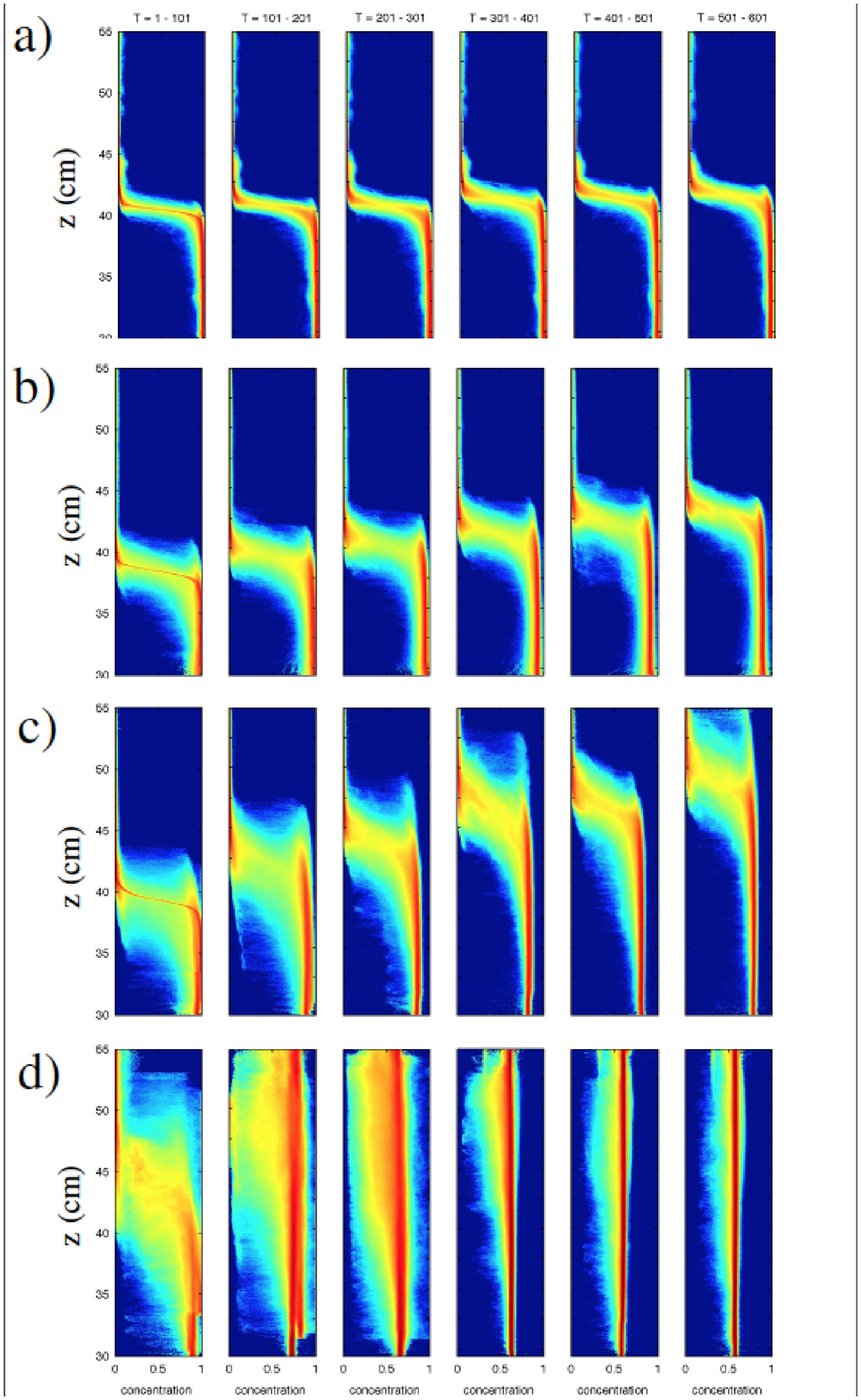}  
\end{center}
\caption{Temporal evolution of the distribution of density levels:  a) EXP4; b) EXP1; c) EXP7; d) EXP6.  For each experiment the density levels are normalised from  0 (light fluid, blue color) to  1 (dense fluid, red color), and a logarithmic scale has been chosen to visualise fluctuations far from the interface.}
\label{fig:pdf}
\end{figure}

Now that we have characterised the properties of the interface elevation, we focus in this subsections on the statistical properties of density fluctuations.  
The temporal evolution of the distribution of density levels $\rho_b(z,\sigma,t)$ is shown Fig. \ref{fig:pdf}.
Each plotted distribution is obtained by building a normalised histogram of density levels for each depth $z$, using  a sequence of 100 images separated by one second,  and each successive plots of  Fig. \ref{fig:pdf} are separated by 100 seconds.
%
%
The maximal density levels decrease slowly with time in the lower layer, with a concomitant increase of the interface height.
An experiment with weak stratification is shown on Fig. \ref{fig:pdf}-d.  
In that case the flow is rapidly fully homogenised, the interface is not well defined,  and the density field does not evolves through a sequence of stationary states.
Around the density interface, the density distributions of Fig. \ref{fig:pdf}  is closed to a double peaked function.
However, the fluctuations of the interface elevation have not be removed to obtain these statistics. 
The doubled peaked function for the density distribution around the interface is therefore mostly due the rapid motion of interfacial waves around a slowly evolving mean interface height. 
The sloshing dynamics of the interface only affects the the density field close to the interface.
Far from the interface, the motion of the density field is not affected by the variations of the interface elevation and  the method used to build the histograms of Fig. \ref{fig:pdf} is relevant to describe density fluctuations.

We see that separating the part of the density statistics due to the wave motion of the interface from the actual turbulent fluctuations is difficult in practice, and we will propose in the following a rudimentary decomposition of the density fields into "waves" and "turbulence" close to the interface.

\subsubsection{The mean vertical profile after removing the effect of large scale interfacial waves}

We present here an experimental test of the statistical mechanics predictions for the mean density profile around the interface. 
As a starting point, we  note that the global distribution in the experimental of density levels is close to a double delta function, for  Richardson numbers from $10$ to $150$, since the ratio of the interface width with the layer depth is much smaller than one.
%
Equilibrium statistical mechanics theory predicts in that case a $\tanh$ shape for the $x$-averaged density profile, see Eq. (\ref{eq:mean_reduced_density}) in subsection \ref{sub:twolevels}.\\
In order to test the statistical mechanics prediction for the density interface, it is necessary to cancel the spurious effect of sloshing dynamics on density statistics close to the interface.
We rebuild for that purpose  the histograms of density levels obtained initially Fig. \ref{fig:pdf}  by considering a frame of reference following the interface height for each image.
The underlying assumption is that the wave motion of the interface is dominated at lowest order by spatial mode with typical length scale larger that the horizontal size of the image.
Assuming that the system is in a quasi-stationary state  on time intervals of $200$ seconds after $t=100$ s for each experiment,   the   $x$-averaged vertical profile of density levels $\overline{b}_{exp}(z)$ is fitted with the function 
\begin{equation}
\overline{b}^{fit}(z)=\frac{\Delta {b} ^{fit}}{2}\left(1- \tanh\left(\frac{z-\overline{h}}{\Delta h^{fit}}\right) \right) \label{eq:fit} .
\end{equation} 
According to the statistical mechanics prediction in Eq.  (\ref{eq:mean_reduced_density}), all experimental vertical mean density profile should collapse 
on the same curve if  the vertical axis is redefined by $z^*=(z-\overline{h})/\Delta h^{fit}$ and if the density is rescaled as  $b^*=b/\Delta b ^{fit}$
Remarkably, all the vertical profile collapse on a curve that is very close to the predicted $tanh$ relation (black curve) on Fig. \ref{fig:exp_mean_fluc}-a.
We note that the fit is better above the interface than below the interface\footnote{As explained in subsection \ref{sub:erfc}, a fit with the error function would be as good as the fit with the $tanh$ function: the two function would be indiscernible on Fig. \ref{eq:mean_reduced_density}-b.}.
A possible reason is  that the tanh-profile corresponds to an equilibrium state for a two level system, while irreversible mixing through turbulent cascade leads to the creation of new intermediate density levels close to the interface.
The effect of irreversible mixing through turbulent cascade is discussed in more detailed in the next subsection.

\subsubsection{Variation of the interface thickness with the Richardson number}

According to the statistical mechanics prediction  Eq. (\ref{eq:mean_reduced_density}), the interface thickness and the Richardson number  $Ri_H$  introduced Eq. (\ref{eq:Ri}) are related through  $\Delta h=2H/Ri_H$. 
We check Fig. \ref{fig:exp_mean_fluc}-b that the observed interface width $\Delta h ^{fit} $ is inversely proportional to the Richardson number $Ri_H$.
However, the coefficient of proportionality is larger than the one predicted by the theory: we observe $\Delta h^{fit} \approx 3 H / Ri_H$.
%
%
A possible reason for this discrepancy may be attributed to our estimate of the rms kinetic energy in the Richardson number  $Ri_H$.
Indeed, $e_c$ were  estimated by considering the rms turbulent kinetic energy measured in a case without  stratification at the height of the interface.
Assuming that the factor $\Delta h^{fit} /\Delta h \approx 3/2$ may be attributed to the modification of turbulent properties due to the stratification, and that the statistical mechanics predictions for the interface height are correct, we  define an effective Richardson number $Ri_H^{fit}$ such that $\Delta h^{fit} = 2H /Ri_H^{fit}$, and we use this Richardson number to estimate the effective energy  $e_c^{fit}=(3/2) H \Delta b /Ri_H^{fit}$. This effective energy can then be used to define the effective temperature close to the interface 
\begin{equation}
\beta^{fit}=\frac{3}{2 e_c^{fit}}=  \frac{1}{\Delta h^{fit} \Delta b^{fit}} \ . \label{eq:betafit} 
\end{equation}
Since the pioneering work of \cite{crapper1974,hopfinger_toly}, the variations of the interface thickness with the Richardson number has remained highly debated, partly because different experimental settings and measurement techniques have been used, partly because different definitions for the interface thickness have been considered. 
\cite{crapper1974,wolanski1975,fernando1985} found that the interface thickness was independent from the Richardson number based not he turbulence length scale ($Ri=\Delta b L_t/(2e_c/3)$).
By contrast, \cite{hopfinger_toly}  distinguished a static thickness $h_s$ measured after stopping mechanical stirring from a dynamical thickness $h_d$ measured in the presence of turbulence. They observed that the static thickness $h_s$ was independent from the Richardson number, while  $(h_d-h_s)/h_s\sim Ri^{-1}$. 
Using PLIF measurements,  \cite{hannoun88}  observed that the mean interface thickness was decreasing with the Richardson number, but with a different scaling ($h_s\sim L_t Ri^{-2}$) and that the interfacial wave amplitude  was decreasing as $h_d\sim L_t Ri^{-1}$ for sufficiently large $Ri$. However, \cite{fernando_hunt2} found   $h_d\sim L_t$ using a similar method (but better spatial resolution) to determine the interface thickness.  
 \cite{hannoun88,fernando_hunt2}  defined the interface thickness $\Delta h(x,t)$ as the height difference between two prescribed iso-density contours (the height difference between the contours $\alpha=0.2$ and $\alpha=0.8$, where $\alpha$ is defined Eq. (\ref{eq:isocontours})). 
Just as in the case of the interface height $h(x,t)$, this method can not be applied to a density field with a strongly corrugated interface, presenting overturning events everywhere.\\ 

Here we have presented two different experimental results supporting a scaling of the interface thickness with the inverse of the Richardson number. 
First, the amplitude of the frequency spectra do collapse on the same curve at high frequency, which is predicted with Phillips theory and the additional assumption that the interface thickness scales as $Ri^{-1}$. 
Second, the fit of the interface shape obtained in a frame of reference following the x-averaged interface elevation  also yields a similar scaling. 
This second test is not fully satisfactory: indeed, in the presence of a perfectly thin interface $\Delta h \approx 0$, with  interfacial waves of wavelength smaller than the windows of observation and wave amplitude scaling as $Ri^{-1}$, our procedure to obtain the interface thickness  would lead to a scaling  $\Delta h \sim Ri^{-1}$.  
We expect that the scaling  $\Delta h\sim Ri^{-1}$ predicted by statistical mechanics is valid for large but moderate Richardson numbers ($Ri\sim 10$), when the interface is permanently breaking, while it is not valid for very high Richardson numbers, when the interface is only  breaking intermittently. 
In this case, the interface acts as a mixing barrier that prevent mixing in physical space, and in phase space. 

\subsubsection{Vertical profile of the variance of density levels}

\begin{figure}
\setlength{\unitlength}{1cm}
\begin{center}
\includegraphics[width=\textwidth]{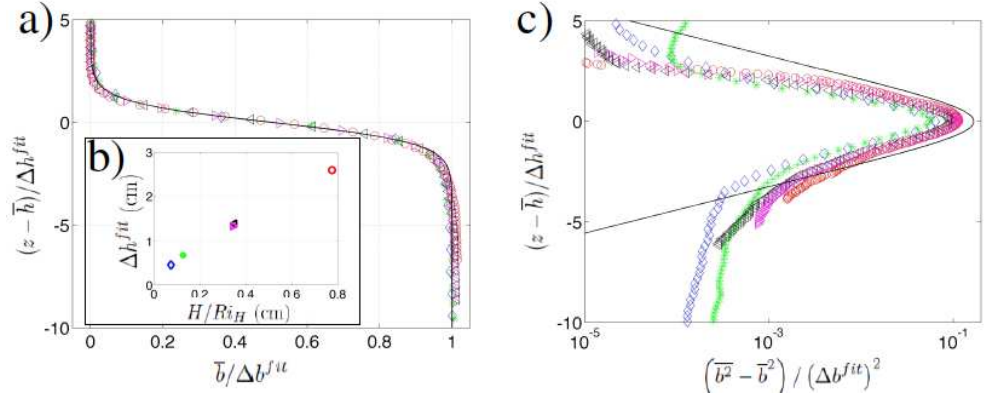}
\end{center}
\caption{
a) Vertical profile of the  $x$-averaged density.
The thin black line is the tanh profile predicted by equilibrium statistical mechanics in the two level case.
Red circle:  EXP7;  black triangle: EXP1; magenta triangle: EXP2;  green star: EXP3; blue diamond: EXP4. 
b) Variation of the interface thickness $\Delta h^{fit}$ with the Richardson number. 
c) Vertical variations of the variance of the density distribution.
The thin black line in the statistical mechanics prediction Eq. (\ref{eq:b_fluct}).
}
\label{fig:exp_mean_fluc}
\end{figure}

Statistical mechanics  predicts not only the mean vertical density profile but also the presence of density fluctuations across the interface. 
These fluctuations can be related to the mean density profile and to the effective temperature of the turbulent flow through Eq. (\ref{eq:beta_fluct}). We assume that the inverse temperature is given by $\beta^{fit}=\left(\Delta h ^{fit} \Delta b^{fit} \right)^{-1}$  defined  Eq. (\ref{eq:betafit}). 
We also assume that the mean density profile 
$\overline{b}(z)$ is well described by the tanh profile defined Eq. (\ref{eq:fit}). Then  Eq. (\ref{eq:beta_fluct}) yields 
\begin{equation}
\frac{\overline{b^{2}}-\overline{b}^{2}}{\left(\Delta b^{fit}\right)^2}= \frac{1}{2}  \left( \cosh\left( \frac{z-\overline{h}}{\Delta h^{fit}}\right) \right)^{-2}  . \label{eq:b_fluct}
\end{equation}
The observed vertical variation of the variance of density fluctuations is plotted in  Fig. \ref{fig:exp_mean_fluc}-c for different experiments (corresponding to different Richardson numbers). 
The vertical coordinate $z^*=(z-\overline{h})/\Delta h^{fit}$ has been rescaled by the interface thickness for each experiment, and we consider the frame of reference following the interface elevation, just as in   Fig.\ref{fig:exp_mean_fluc}-a. The variance  of density fluctuations is rescaled by $\left(\Delta b^{fit}\right)^2$. 

The thin black line is the statistical mechanics prediction given by Eq. (\ref{eq:b_fluct}), 
This theoretical prediction is  qualitatively correct sufficiently close to the interface.
Far above the interface, the statistical mechanics theory  overestimates the fluctuations : these fluctuations are absent in the upper layer  since the turbulence is located in the lower layer and around the interface.
Far below the interface, within the mixed layer, equilibrium theory underestimate the density fluctuations: filaments of light fluids entrained in the mixed layer are not stirred as much as would be predicted by the equilibrium theory: the density field is strongly out of equilibrium in this region. 
%


Although the observed vertical profiles of density fluctuations close to the interface are close to the predicted vertical profile, their amplitude remains smaller than the  amplitude predicted by the equilibrium theory.
The main reason for this discrepancy is that the theory does not take into account irreversible mixing of density, which tends to decrease the density fluctuations.
In addition, we clearly see on Fig. (\ref{fig:pdf}) that the density distributions close to the interface contains a continuum of density levels between the extremal values.
The presence of these intermediate density levels which were not initially present in the two layer density field are also evidence for irreversible mixing by turbulent cascade.
This effect can not be captured in the framework of the simple two level system. 
Modelling the combined effect of irreversible mixing (through turbulence cascade) and of the relaxation toward equilibrium has partly been addressed by \cite{venaillesommeria2011} and will be the object of future work\\

To conclude, one can distinguish two regions for the density field:
 i/ The region close to the density interface, which may be interpreted as an equilibrium state once the effect of interfacial waves and interface increase due to entrainment are removed. 
The theory predict correctly the mean vertical density profile, but overestimate the fluctuations, and the observed density distribution is different from the initially postulated two level distribution. 
 ii/ The region far from the interface ($z\gg \Delta h$), which is strongly out of equilibrium since the observed variance of density fluctuations is much larger than the one predicted by the equilibrium theory.
The aim of the next section is to  describe in more details the properties of density fluctuations in this out-of equilibrium region. 

\subsection{Exponential shape of the tracer distribution far from the interface}

\begin{figure}
\setlength{\unitlength}{1cm}
\begin{center}
\includegraphics[height=7cm]{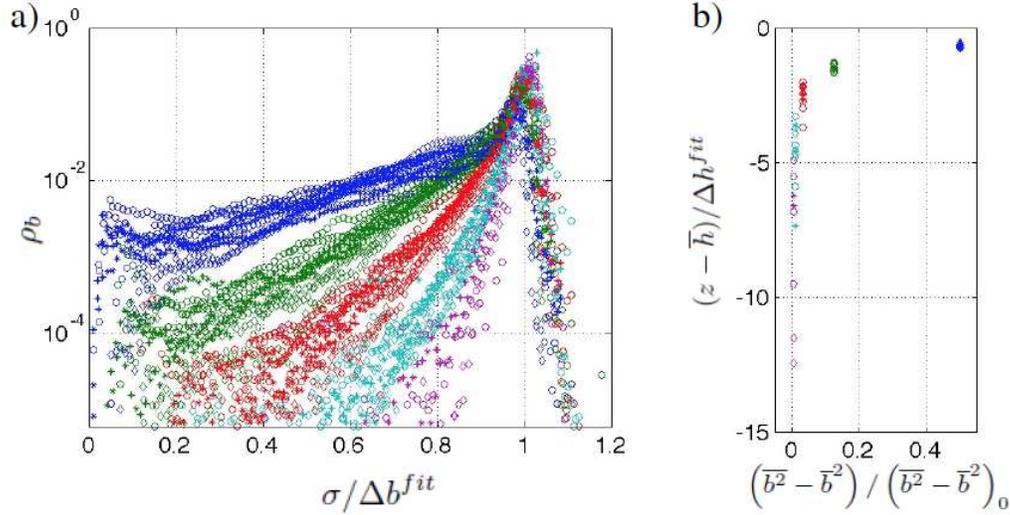}
\end{center}
\caption{a) Probability distribution function of the density at increasing distances from the interface.   Each symbol corresponds to a different experiment ($o$: EXP7, $\diamond$: EXP2,  $*$: EXP4). For each experiment,  5 different depth (associated with 5 different colors on the plot) are  considered. Those 5 depth are chosen such that the density variances of the pdf normalised by the density variance at the interface are  the same for a given color (the relation between depth and the prescribed values of density variance is shown on panel b). 
For each experiment, the density levels on the $x$ axis are normalised by  the maximum value of the x-averaged density, denoted  $\Delta b^{fit}$. 
%
b) Variation of the  height (vertical axis) associated with  prescribed values of relative density variance  ($\left(\overline{b^2}- \overline{b}^2\right)/\left(\overline{b^2}- \overline{b}^2\right)_0$ on the horizontal axis).
}
\label{fig:pdf_variance}
\end{figure}

 %
Sufficiently far from the interface, the fluid motion is not influenced by the motion induced by the interfacial waves, and the distribution of density levels observed in Fig. \ref{fig:pdf} (obtained without change of reference frame) can directly be interpreted as turbulent fluctuations.
We see on Fig. \ref{fig:pdf_variance}-a   the distribution of  density levels plotted  for each experiment at five different  depth corresponding to five prescribed  values of the  density  variance (relative to the density variance at the interface for each experiment).   
The density levels on the horizontal axis are normalised by $\Delta b^{fit}$. According to Eq. (\ref{eq:fit}), $\Delta b^{fit}$  is the maximal value of the $x$-averaged vertical  profile of density for each experiment.
Since the initial density jump is $\Delta b_0>\Delta b^{fit}$,   density levels larger than $\Delta b^{fit}$ can be observed.
The depth  and normalised density variance corresponding to each plotted density distribution are shown Fig. \ref{fig:pdf_variance}-b.
We see on Fig. (\ref{fig:pdf})   that  a given value of the density variance  is associated with a well defined depth when sufficiently close to the interface, but that the depth 
associated with a given value of the variance are more scattered far from the interface.

Strikingly, the density distributions  of Fig. \ref{fig:pdf_variance}-a do collapse qualitatively well, and are characterised by exponential tails, with an e-folding depth that decreases at increasing distances from the interface.\\

Exponential tails in the  distribution of a tracer in turbulent flow have been previously reported either in the case  of an isolated source discharging the tracer into an infinite (unconfined) medium \citep{duplat2010}, or in the case a confined medium with large scale inhomogeneities due the  injection of the tracer at the boundaries.
This is for instance the case in convection experiments, where exponential tails in the temperature distribution have been reported for very high Reynolds numbers \citep{castaing1989}.
In these convection experiments, one can consider that the temperature in the bulk is statistically homogeneous.
The statistically steady distributions result from a competition between turbulent cascade that tends to dissipate temperature fluctuations and the tracer fluxes at the boundaries that inject fluctuations in the bulk \citep{pumir1991}.
In this approach, the density is considered as a passive tracer
As far as the distribution of density levels is concerned, the mixed layer  in the present stably stratified experiment is analogous to the mixed layer in the convection experiment at very high Reynolds number.
The only difference is that  the source of density fluctuations in the stably stratified experiment comes only from the interface at the top of the mixed layer, while the injection of density fluctuations comes from both the upper and the lower layer in convection experiments.
This asymmetry in the injection of density fluctuations explains why only one tail of the density distribution present an exponential shape in the case of the stably stratified experiment.

\section{Conclusion}

We have proposed a statistical mechanics  interpretation of the formation of sharp but highly corrugated density interface between region of homogeneous density  in the presence of turbulence, building upon previous work by \cite{tabak2004}, which generalises the Miller-Robert-Sommeria approach for the vorticity in two-dimensional turbulence to the density  in three dimensional stratified turbulence. The statistics of the density field is predicted as  the most probable outcome of turbulent stirring. An effective "heat" bath is provided by the turbulent velocity field. The temperature of this "heat bath" is proportional to the turbulent eddy kinetic energy. 
In the case of a system initially composed of two homogeneous layers with a stable density interface, the theory predicts a  tanh-shape for the mean vertical density profile.
This equilibrium density  profile is interpreted as the result of a competition between turbulent transport that tends to smooth out the interface and buoyancy forces that tend to sharpen the interface by the sorting of the fluid elements by density.
For large Richardson numbers, buoyancy takes over turbulent transport and the interface is thin.
More precisely, the interface thickness is inversely proportional to the Richardson number. 

%
%
%
%

The equilibrium theory alone  cannot describe entrainment across the interface, which would eventually lead to complete homogenisation of the density field. In the  case of the experiments presented in this paper, turbulence generation is limited to the lower layer. The interface is progressively drifting upward by the entrainment of fresh fluid, and the lower layer density decreases due to this mixing process.
Entrainment is related to irreversible mixing of density levels  through turbulent cascade, a process which changes the global distribution of density levels.
We propose a simplified model which keeps a two level system at statistical equilibrium, accounting for the effect of entrainment by a progressive decrease of the lower layer density.

The advantage of the statistical mechanics  approach over previous models is that it provides a prediction for finite interface thickness in the presence of a source of turbulence without assumption about turbulent diffusivity.
In addition, it does not rely on a particular mechanism at stake close to the interface (wave breaking, Kelvin-Helmholtz instability,...) that may depend on the Richardson number \citep{fernando_hunt2}.
The only assumption is  that the system sufficiently explores the phase space. 

In order to test the equilibrium theory, we assumed the existence of a statistically stationary state, and obtained statistics of the density field close to the interface using a reference frame following the horizontally averaged interface height.
This method allowed us to get rid of the out-of equilibrium effects of the entrainment and of the gravest spatial modes of the  interfacial gravity waves (which were found to dominate the temporal fluctuations of the interface elevation).
We found that the  shape of the observed mean vertical density profile  is well fitted by a hyperbolic tangent function, as predicted by the equilibrium theory in the case of a two level system. Furthermore the interface thickness was found to be inversely proportional to the Richardson number between Ri $\sim 10$ and Ri $\sim 100$, as expected form the theory.
We do not know whether these predictions would remain valid for higher Richardson numbers (still for a turbulent flow).
Indeed, the interface then becomes locally sharper than the resolution of our measurements, with spatially and temporally intermittent mixing events across this interface, see also \cite{fernando_hunt2}. 
The statistical theory also provides a good prediction for the vertical profile of the rms density fluctuation variance at the interface. The value of this variance is however smaller than predicted by theory. The discrepancy is probably due to the dissipation of density fluctuations by the turbulent cascade.

Below the interface, the measured density fluctuations are by contrast much stronger than the statistical equilibrium prediction, which is very low in relation with the quasi-uniform mean density, according to Eq. (\ref{eq:eq_statmom}). These density fluctuations are strongly out-of equilibrium: they are transported from the interface by turbulent transport processes.
In this region, we observed exponential tails in the probability distribution function of density levels.
Such exponential tails can be attributed to isolated filaments that are entrained from the interface and then stirred through turbulent cascade in the lower layer. The  density then behaves as a passive scalar, and the shape of its pdf results from a competition between turbulent cascade and the injection of light density filaments entrained from the interface.
This situation is analogous to convection experiment at high Reynolds numbers, where exponential tails in the density pdf have been also previously reported.\\

Density fluctuations discussed above correspond to rather small scales, for which turbulent motion prevails. At larger scale the interface fluctuates as internal waves, for which we were able to check the dispersion relation. We observe that the  spatial and temporal spectra of these waves can be interpreted by a theory due to \cite{phillips77} which states that the wave amplitude at each frequency is such that the induced flow is close to criticality for shear instability at the interface. In the range of frequency  much larger than the gravest mode, and much smaller than the buoyancy frequency characterising the density gradient inside the interface, this theory predicts a  $-3$ slope for the interface elevation frequency power spectrum. We confirm this result, which also agrees with previous observations by \cite{hannoun88}.
The value of the predicted spectral energy depends on the interface thickness as an input. We observed that the amplitude of the waves in this range of frequency is independent from the Richardson number, which corresponds to the statistical equilibrium prediction of an interface thickness varying as the inverse of the Richardson number. 
This scaling is different than the one obtained by  \cite{hannoun88}, and we do not know whether it would still be valid for higher Richardson number (beyond 100), as explained above. Note finally that the Phillipp's approach is limited to frequencies for which the turbulent velocities are negligible with respect to the potential flow associated with the variations of the interface elevation. A more detailed study of the  coupling between interfacial waves and turbulence has been addressed by \cite{fernando_hunt1,fernando_hunt2}.\\
We focused in this paper to the visualisation of the density field using PLIF technique with index matching. 
The only PIV measurement presented in this paper were performed in a case without stratification. 
The underlying assumption was that turbulence properties were not much affected by the presence of stratification (below the interface). 
We chose to use PLIF only in order to obtain quantitative  measurements of  turbulent fluctuations in the density field (the presence of particle used the PIV measurements alters the quality of the visualisation).
However, we hope to address in future for a more detailed study of the buoyancy fluxes close to the interface using Simultaneous PIV and PLIF measurement, which has recently been performed in the context of entrainment in gravity currents \citep{odier2014}.\\

To conclude, equilibrium statistical mechanics allows us to interpret qualitatively the formation of thin but corrugated and turbulent  density interface between  regions of homogeneous density.  The approach is limited by out-of equilibrium effect such as irreversible mixing and interfacial wave excitation. 
We believe that combining these different phenomenological approaches will lead to fruitful models in the context of  turbulent mixing across a density interface, or more generally in any stratified turbulence problem where density fluctuations play an important role, as for intense in gravity currents \citep{odier2009,odier2012}. It would be also interesting to check the relevance of equilibrium statistical mechanics at the interface  of two immiscible fluids. On the one hand, this would avoid the issue with irreversible mixing through turbulence cascade, but on the other hand, surface tension effects may be influential.

\section{Appendix A: Equipartition and mixing efficiency of $0.25$ in the low energy limit}

Eq. (\ref{eq:beta_fluct}) allows to derive an interesting side result concerning energy equipartition in the low energy limit. 
This result may in turn be used to predict a mixing efficiency coefficient, which is a measure of the  fraction of the  energy injected in a stratified fluid that is actually used to irreversibly increase the potential energy of the flow. 

 Let us consider a stratified fluid initially at rest, characterised by its density profile $b_{bg}(z)$ (the index ``bg'' stands for ``background'').
The potential energy of this state if denoted $E_{bg,p}\equiv\int_0^H\mathrm{d} z\ b z$. 
Let us then assume that the fluid is isolated, that a given amount of energy $\Delta E$ is injected in the system, and that a statistical equilibrium is reached on a time scale much shorter than the time scale for viscous dissipation and for irreversible mixing of density. Let us call $E_c$ and $E_p$ the kinetic and the potential energy of the equilibrium state. The available potential energy of the equilibrium state is
\begin{equation}
\Delta E_p \equiv \int\int\int  \mathrm{d} x \mathrm{d} y \mathrm{d} z  \left( b -b_{bg} \right) z = \int \mathrm{d} z  \left( \overline{b} -b_{bg} \right). \label{eq:ape_def} 
\end{equation}
The second equality is obtained by noting that the equilibrium state is statistically invariant on the horizontal, and that  the horizontal integral amounts to an ensemble average. 
For the equilibrium state, the potential energy $\Delta E_p$ may be qualified as available since one  recover $\overline{b}=b_{bg}$ just by setting the effective temperature to 0 ($\beta=+\infty$).
Since each fluid particle conserves its density in the absence of dissipation, a  fluid particle of density $b$ at height $z$ and time $t$ can be seen as a fluid particle initially at height $z_{bg}(b)$, displaced  by  $\xi(x,y,z)=z-z_{bg}(b(x,y,z))$ in the vertical direction. 
Let us assume that this displacement $\xi $ is sufficiently small, which is ensured by considering a low energy limit. 
At lowest order, the density fluctuation defined by $b'(x,y,z)=b(x,y,z)-b_{bg}(z)$ is proportional to the vertical fluid particle displacement: $b'=\xi \partial_z b_{bg}$, and the available potential energy can be expressed (still at lowest order) as 
 \begin{equation}
 \Delta E_p = \frac{1}{2}\int\int\int  \mathrm{d} x \mathrm{d} y \mathrm{d} z \frac{b'^2}{\partial_z b_{bg}} =\frac{1}{2} \int \mathrm{d} z \frac{\overline{b^2}-\overline{b}^2}{\partial_z b_{bg}}.   \label{eq:ape_appbis}
\end{equation}
Again, the second equality is obtained by noting  the horizontal integration amounts to an ensemble average. 
Injecting then Eq. (\ref{eq:beta_fluct}) in Eq. (\ref{eq:ape_appbis})  gives
\begin{equation}
\Delta E_p = \frac{1}{2\beta}\int \mathrm{d}z \frac{\partial_z \overline{b}}{\partial_z b_{bg}} .\label{eq:ratio}
\end{equation}
In the low energy limit, we get at lowest order  $\partial_z \overline{b}\approx \partial_z b_{bg} $, which  yields $\Delta E_p = V/(2\beta)$, where $V$ is the volume where the flow takes place. 
The inverse temperature $\beta$ is related to the total kinetic energy $\Delta E_c = V e_c $ through Eq. (\ref{eq:energy-beta}), which yields $\Delta E_p =\Delta E_c/3$. 
This expresses equipartition of the energy between the available potential energy and the three degrees of freedom of the kinetic energy.

Finally, the ratio of the available potential energy with the total energy injected in the system is 
\begin{equation}
\eta\equiv \frac{ \Delta E_p}{\Delta E_p+\Delta E_c}=\frac{1}{4}.\label{eq:def_eta}
\end{equation} 
Let us now assume that once the equilibrium state is reached, the density fluctuations are  smoothed out on each horizontal plane due to the combined effect of direct turbulent cascade and molecular diffusivity. 
Let us also assume that the rate of kinetic energy dissipation is equal to the rate of dissipation for the density variance at each height. 
These hypothesis ensure that the rhs of Eq. (\ref{eq:beta_fluct}) remains constant through the flow evolution,  and that the profile $\overline{b}$ remains the equilibrium state throughout the flow evolution.
At sufficiently large time, once the fluctuations around  $\overline{b}$  are irreversibly mixed, the flow is at rest and this $\overline{b}$ becomes the new background density profile. 
The increase of potential energy  $\Delta E_p$ defined by  Eq. (\ref{eq:ape_def}) accounts therefore for the irreversible increase of potential energy 
\footnote{Note that  Eq. (\ref{eq:ape_appbis}) is derived from  Eq. (\ref{eq:ape_def}) by assuming that each fluid particle conserves its density; these equations are no more equivalent once the fluctuations of density have been smoothed out at each height $z$ (in which case the available potential energy vanish).
However, the result  $\Delta E_p=\Delta E_c /3$ remains valid since it is obtained at equilibrium.}. 
The mixing efficiency defined as the irreversible increase of potential energy normalised by the total energy injected in the system is then simply given by Eq. (\ref{eq:def_eta}).

To conclude, a mixing efficiency coefficient of $1/4$ can be interpreted as a result of energy equipartition at equilibrium under the following assumption:  i/ a low energy limit ii/  a  time scale for the  dissipation of horizontal density fluctuations much larger than the time scale to reach the equilibrium state  iii/ a rate of dissipation of the variance of density fluctuation at each height equal to the rate of dissipation of the kinetic energy.
If these conditions are not fulfilled, then the mixing efficiency should be smaller than $0.25$.

Mixing efficiency coefficients between 0.2 and 0.3 are widely used in modelling context.  However, experiments and simulations and observations seem to show that there is no universal mixing efficiency in stably stratified turbulence \cite{peltier2003,ivey2008}. Our result suggest that the value $0.25$ may be interpreted as a limit case in the framework of equilibrium statistical mechanics.

\section{Appendix B : Turbulence properties in the homogeneous case}

We performed one experiment without stratification in order to have a reference flow that we analysed using PIV measurements.
This allowed us to obtain some characteristics of the flow in the turbulent layer only, assuming that these flow properties would not be much different in the presence of stratification provided that one consider the dynamics sufficiently below the interface.
In particular, this methods allows to estimate the energy flux due to the grid forcing, and the decay of the turbulence strength with altitude.


Let us call $L_t(z)$ the integral length scale of turbulence.
Sufficiently close to the source, the integral length scale of turbulence if given by the grid mesh of the oscillating grid $L_t=L_f$.
Sufficiently far from the source, the only length scale of the problem is the distance from the source and $L_t \sim  z$. 
The temporal evolution of the kinetic energy may be modelled as   
\begin{equation}
\partial_t e_c=  \partial_z \left( \nu_t \partial_z e_c \right)- \frac{c_d}{L_t} e_c^{3/2},\label{eq:ec_dyn}
\end{equation}
where the effect of turbulence is modelled as  effective viscosity, with a turbulent energy flux  
\begin{equation}
F_{e_c}=- \nu_t  \partial_z e_c, \quad \nu_t =a e_c^{1/2} L_t  \ .\label{eq:turb_flux}
\end{equation}
The sink of energy is given by a Kolmogorov dissipation term.
Sufficiently close to the grid, $L_t=L_f$ is a constant and stationary  kinetic energy profile is exponential: 
\begin{equation}
e_c= e_c^0 \exp\left(-\frac{z}{L_t} \right) ,\ L_t=\left( \frac{a}{c_d} \right)^{1/2}L \label{eq:ec_stat} 
\end{equation}
with $e_c^0$ the kinetic energy at the source located in $z=0$.
Sufficiently  far from the source $L_t \sim z$ and stationary kinetic energy profile is given by a power law  $e_c \sim z^{-2}$, see e.g.

\cite{hopfinger_toly}.



%
%
%
%
%
%
%
%
%
%
%
%
%
%
%
%
%
%
%
%
%
%
%
%
%
%
%
%
%

\bibliographystyle{jfm}
\bibliography{twolayers}

\begin{thebibliography}{47}
\expandafter\ifx\csname natexlab\endcsname\relax\def\natexlab#1{#1}\fi

\bibitem[Balmforth {\em et~al.\/}(1998)Balmforth, Llewellyn~Smith \&
  Young]{balmforth1998}
{\sc Balmforth, N.J., Llewellyn~Smith, S.G. \& Young, W.R.} 1998 Dynamics of
  interfaces and layers in a stratified turbulent fluid. {\em Journal of Fluid
  Mechanics\/} {\bf 355}, 329--358.

\bibitem[Bouchet \& Venaille(2012)]{bouchet2012}
{\sc Bouchet, F. \& Venaille, A.} 2012 Statistical mechanics of two-dimensional
  and geophysical flows. {\em Physics reports\/} {\bf 515}~(5), 227--295.

\bibitem[Carruthers \& Hunt(1986)]{carruthers1986}
{\sc Carruthers, D.J. \& Hunt, J.C.R.} 1986 Velocity fluctuations near an
  interface between a turbulent region and a stably stratified layer. {\em
  Journal of Fluid Mechanics\/} {\bf 165}, 475--501.

\bibitem[Castaing {\em et~al.\/}(1989)Castaing, Gunaratne, Heslot, Kadanoff,
  Libchaber, Thomae, Wu, Zaleski \& Zanetti]{castaing1989}
{\sc Castaing, B., Gunaratne, G., Heslot, F., Kadanoff, L., Libchaber, A.,
  Thomae, S., Wu, X.-Z., Zaleski, S. \& Zanetti, G.} 1989 Scaling of hard
  thermal turbulence in rayleigh-b{\'e}nard convection. {\em Journal of Fluid
  Mechanics\/} {\bf 204}, 1--30.

\bibitem[Crapper \& Linden(1974)]{crapper1974}
{\sc Crapper, P.F. \& Linden, P.F.} 1974 The structure of turbulent density
  interfaces. {\em Journal of Fluid Mechanics\/} {\bf 65}~(01), 45--63.

\bibitem[Duplat {\em et~al.\/}(2010)Duplat, Innocenti \&
  Villermaux]{duplat2010}
{\sc Duplat, J., Innocenti, C. \& Villermaux, E.} 2010 A nonsequential
  turbulent mixing process. {\em Physics of Fluids (1994-present)\/} {\bf
  22}~(3), 035104.

\bibitem[{E} \& {Hopfinger}(1986)]{E_hopfinger86}
{\sc {E}, X. \& {Hopfinger}, E.~J.} 1986 {On mixing across an interface in
  stably stratified fluid}. {\em Journal of Fluid Mechanics\/} {\bf 166},
  227--244.

\bibitem[Fernando \& Long(1985)]{fernando1985}
{\sc Fernando, H.J.S. \& Long, R.R.} 1985 On the nature of the entrainment
  interface of a two-layer fluid subjected to zero-mean-shear turbulence. {\em
  Journal of Fluid Mechanics\/} {\bf 151}, 21--53.

\bibitem[{Fernando}(1991)]{fernando91}
{\sc {Fernando}, H.~J.~S.} 1991 {Turbulent mixing in stratified fluids}. {\em
  Annual Review of Fluid Mechanics\/} {\bf 23}, 455--493.

\bibitem[{Fernando} \& {Hunt}(1997)]{fernando_hunt1}
{\sc {Fernando}, H.~J.~S. \& {Hunt}, J.~C.~R.} 1997 {Turbulence, waves and
  mixing at shear-free density interfaces. Part 1. A theoretical model}. {\em
  Journal of Fluid Mechanics\/} {\bf 347}, 197--234.

\bibitem[{Guyez} {\em et~al.\/}(2007){Guyez}, {Flor} \&
  {Hopfinger}]{guyez_flor}
{\sc {Guyez}, E., {Flor}, J.-B. \& {Hopfinger}, E.~J.} 2007 {Turbulent mixing
  at a stable density interface: the variation of the buoyancy flux gradient
  relation}. {\em Journal of Fluid Mechanics\/} {\bf 577}, 127.

\bibitem[{Hannoun} \& {List}(1988)]{hannoun88}
{\sc {Hannoun}, I.~A. \& {List}, E.~J.} 1988 {Turbulent mixing at a shear-free
  density interface}. {\em Journal of Fluid Mechanics\/} {\bf 189}, 211--234.

\bibitem[{Hopfinger} \& {Toly}(1976)]{hopfinger_toly}
{\sc {Hopfinger}, E.~J. \& {Toly}, J.-A.} 1976 {Spatially decaying turbulence
  and its relation to mixing across density interfaces}. {\em Journal of Fluid
  Mechanics\/} {\bf 78}, 155--175.

\bibitem[Ivey {\em et~al.\/}(2008)Ivey, Winters \& Koseff]{ivey2008}
{\sc Ivey, GN, Winters, KB \& Koseff, JR} 2008 Density stratification,
  turbulence, but how much mixing? {\em Annual Review of Fluid Mechanics\/}
  {\bf 40}~(1), 169.

\bibitem[Linden(1973)]{linden1973}
{\sc Linden, P.F.} 1973 The interaction of a vortex ring with a sharp density
  interface: a model for turbulent entrainment. {\em Journal of Fluid
  Mechanics\/} {\bf 60}~(03), 467--480.

\bibitem[Linden(1979)]{linden1979}
{\sc Linden, P.F.} 1979 Mixing in stratified fluids. {\em Geophysical \&
  Astrophysical Fluid Dynamics\/} {\bf 13}~(1), 3--23.

\bibitem[Linden(1980)]{linden1980}
{\sc Linden, P.F.} 1980 Mixing across a density interface produced by grid
  turbulence. {\em Journal of Fluid Mechanics\/} {\bf 100}~(04), 691--703.

\bibitem[Lucarini {\em et~al.\/}(2013)Lucarini, Blender, Herbert, Pascale \&
  Wouters]{lucarini2014}
{\sc Lucarini, V., Blender, R., Herbert, C., Pascale, S. \& Wouters, J.} 2013
  Mathematical and physical ideas for climate science. {\em arXiv preprint
  arXiv:1311.1190\/} .

\bibitem[Majda \& Wang(2006)]{majda2006}
{\sc Majda, A. \& Wang, X.} 2006 {\em Nonlinear dynamics and statistical
  theories for basic geophysical flows\/}. Cambridge University Press.

\bibitem[Mcdougall(1979)]{mcdougall1979}
{\sc Mcdougall, Trevor~J} 1979 Measurements of turbulence in a zero-mean-shear
  mixed layer. {\em Journal of Fluid Mechanics\/} {\bf 94}~(03), 409--431.

\bibitem[{McGrath} {\em et~al.\/}(1997){McGrath}, {Fernando} \&
  {Hunt}]{fernando_hunt2}
{\sc {McGrath}, J.~L., {Fernando}, H.~J.~S. \& {Hunt}, J.~C.~R.} 1997
  {Turbulence, waves and mixing at shear-free density interfaces. Part 2.
  Laboratory experiments}. {\em Journal of Fluid Mechanics\/} {\bf 347},
  235--261.

\bibitem[Miles(1961)]{miles1961}
{\sc Miles, J.W.} 1961 On the stability of heterogeneous shear flows. {\em
  Journal of Fluid Mechanics\/} {\bf 10}~(04), 496--508.

\bibitem[Miller(1990)]{miller1990}
{\sc Miller, J.} 1990 Statistical mechanics of euler equations in two
  dimensions. {\em Physical review letters\/} {\bf 65}~(17), 2137.

\bibitem[Mory(1991)]{mory1991}
{\sc Mory, M.} 1991 A model of turbulent mixing across a density interface
  including the effect of rotation. {\em Journal of Fluid Mechanics\/} {\bf
  223}, 193--207.

\bibitem[Odier {\em et~al.\/}(2012)Odier, Chen \& Ecke]{odier2012}
{\sc Odier, P., Chen, J. \& Ecke, R.E.} 2012 Understanding and modeling
  turbulent fluxes and entrainment in a gravity current. {\em Physica D:
  Nonlinear Phenomena\/} {\bf 241}~(3), 260--268.

\bibitem[Odier {\em et~al.\/}(2014)Odier, Chen \& Ecke]{odier2014}
{\sc Odier, P., Chen, J. \& Ecke, R.~E.} 2014 Entrainment and mixing in a
  laboratory model of oceanic overflow. {\em Journal of Fluid Mechanics\/} {\bf
  746}, 498--535.

\bibitem[Odier {\em et~al.\/}(2009)Odier, Chen, Rivera \& Ecke]{odier2009}
{\sc Odier, P., Chen, J., Rivera, M.~K. \& Ecke, R.~E.} 2009 Fluid mixing in
  stratified gravity currents: The prandtl mixing length. {\em Physical review
  letters\/} {\bf 102}~(13), 134504.

\bibitem[Park \& Gnanadeskian(1994)]{park1994}
{\sc Park, Y.-G.and~Whitehead, J.A. \& Gnanadeskian, A.} 1994 Turbulent mixing
  in stratified fluids: layer formation and energetics. {\em Journal of Fluid
  Mechanics\/} {\bf 279}, 279--311.

\bibitem[Peltier \& Caulfield(2003)]{peltier2003}
{\sc Peltier, WR \& Caulfield, CP} 2003 Mixing efficiency in stratified shear
  flows. {\em Annual review of fluid mechanics\/} {\bf 35}~(1), 135--167.

\bibitem[Phillips(1972)]{phillips}
{\sc Phillips, O.M.} 1972 Turbulence in a strongly stratified fluid: is it
  unstable? In {\em Deep Sea Research and Oceanographic Abstracts\/}, ,
  vol.~19, pp. 79--81. Elsevier.

\bibitem[Phillips(1977)]{phillips77}
{\sc Phillips, O.M.} 1977 The dynamic of upper ocean .

\bibitem[Posmentier(1977)]{posmentier}
{\sc Posmentier, E.S.} 1977 The generation of salinity finestructure by
  vertical diffusion. {\em Journal of Physical Oceanography\/} {\bf 7}~(2),
  298--300.

\bibitem[Pumir {\em et~al.\/}(1991)Pumir, Shraiman \& Siggia]{pumir1991}
{\sc Pumir, A., Shraiman, B.~I. \& Siggia, E.~D.} 1991 Exponential tails and
  random advection. {\em Physical review letters\/} {\bf 66}, 2984--2987.

\bibitem[Renaud {\em et~al.\/}(2014)Renaud, Venaille \& Bouchet]{renaud2014}
{\sc Renaud, A., Venaille, A. \& Bouchet, F.} 2014 Equilibrium states and
  energy partition of the shallow water model. {\em preprint\/} {\bf 00}~(00),
  000.

\bibitem[Robert \& Sommeria(1991)]{robert1991}
{\sc Robert, R. \& Sommeria, J.} 1991 Statistical equilibrium states for
  two-dimensional flows. {\em Journal of Fluid Mechanics\/} {\bf 229},
  291--310.

\bibitem[{Rouse, H.} \& {Dodu, J.}(1955)]{rouse_dodu}
{\sc {Rouse, H.} \& {Dodu, J.}} 1955 Turbulent diffusion across a density
  discontinuity. {\em La Houille Blanche\/} {\bf 4}, 522--532.

\bibitem[Ruddick {\em et~al.\/}(1989)Ruddick, McDougall \& Turner]{ruddick1989}
{\sc Ruddick, B.R., McDougall, T.J. \& Turner, J.S.} 1989 The formation of
  layers in a uniformly stirred density gradient. {\em Deep Sea Research Part
  A. Oceanographic Research Papers\/} {\bf 36}~(4), 597--609.

\bibitem[Salmon(1998)]{salmon1998}
{\sc Salmon, R.} 1998 {\em Lectures on geophysical fluid dynamics\/}, , vol.
  378. Oxford University Press Oxford.

\bibitem[Schmitt(1994)]{schmitt1994}
{\sc Schmitt, R.W.} 1994 Double diffusion in oceanography. {\em Annual Review
  of Fluid Mechanics\/} {\bf 26}~(1), 255--285.

\bibitem[Simpson \& Woods(1970)]{simpson1970}
{\sc Simpson, J.H. \& Woods, J.D.} 1970 Temperature microstructure in a fresh
  water thermocline. {\em Nature\/} .

\bibitem[Sommeria(2001)]{sommeria2001}
{\sc Sommeria, J.} 2001 Two-dimensional turbulence. In {\em New trends in
  turbulence Turbulence: nouveaux aspects\/}, pp. 385--447. Springer.

\bibitem[Sullivan(1972)]{sullivan1972}
{\sc Sullivan, P.J.} 1972 The penetration of a density interface by heavy
  vortex rings. {\em Water, Air, and Soil Pollution\/} {\bf 1}~(3), 322--336.

\bibitem[Tabak \& Tal(2004)]{tabak2004}
{\sc Tabak, E.G. \& Tal, F.A.} 2004 Mixing in simple models for turbulent
  diffusion. {\em Communications on pure and applied mathematics\/} {\bf
  57}~(5), 563--589.

\bibitem[{Turner}(1968)]{turner68}
{\sc {Turner}, J.~S.} 1968 {The influence of molecular diffusivity on turbulent
  entrainment across a density interface}. {\em Journal of Fluid Mechanics\/}
  {\bf 33}, 639--656.

\bibitem[Venaille \& Sommeria(2010)]{venaillesommeria2011}
{\sc Venaille, A. \& Sommeria, J.} 2010 Modeling mixing in two-dimensional
  turbulence and stratified fluids. In {\em IUTAM Symposium on Turbulence in
  the Atmosphere and Oceans: Proceedings of the IUTAM Symposium on Turbulence
  in the Atmosphere and Oceans, Cambridge, UK, December 8-12, 2008\/}, ,
  vol.~28, p. 155. Springer.

\bibitem[Whitehead \& Stevenson(2007)]{whitehead2007}
{\sc Whitehead, J.A. \& Stevenson, I.} 2007 Turbulent mixing of two-layer
  stratified fluid. {\em Physics of Fluids (1994-present)\/} {\bf 19}~(12),
  125104.

\bibitem[Wolanski(1975)]{wolanski1975}
{\sc Wolanski, E.J .and~Brush, L.M.} 1975 Turbulent entrainment across stable
  density step structures. {\em Tellus\/} {\bf 27}~(3), 259--268.

\end{thebibliography}

\end{document}